\documentclass[amsmath,amssymb,showpacs,showkeywords,twocolumn]{revtex4}
\usepackage[dvips]{graphicx,color}
\usepackage{times}
\usepackage{mathrsfs}
\usepackage{amsmath}
\usepackage{graphicx}
\usepackage{epsfig}
\usepackage{dcolumn}
\usepackage{bm}
\begin{document}
\title{Spin transport and Spin Tunnelling Magneto-Resistance (STMR) of F$|$NCSC$|$F spin valve}
\author{Saumen Acharjee\footnote{saumenacharjee@gmail.com} and Umananda Dev 
Goswami\footnote{umananda2@gmail.com}}
\affiliation{Department of Physics, Dibrugarh University, Dibrugarh 786 004, 
Assam, India}

\begin{abstract}
In this work, we study the spin transport at the 
Ferromagnet$|$Noncentrosymmetric Superconductor (F$|$NCSC) junction of a 
Ferromagnet$|$Noncentrosymmetric Superconductor$|$Ferromagnet (F$|$NCSC$|$F) 
spin valve. We investigate the Tunnelling Spin-Conductance (TSC), spin current 
and Spin Tunnelling Magneto-Resistance (STMR), and their dependence on 
various important parameters like Rashba Spin-Orbit Coupling (RSOC), 
strength and orientation of magnetization, an external in-plane magnetic 
field, barrier strength and a significant Fermi Wavevector Mismatch (FWM) at 
the ferromagnetic and superconducting regions. The study has been carried 
out for different singlet-triplet mixing of the NCSC gap parameter.
We develop Bogoliubov-de Gennes (BdG) Hamiltonian and use the extended 
Blonder - Tinkham - Klapwijk (BTK) approach along with the scattering matrix 
formalism to calculate the scattering coefficients. Our results strongly 
suggest that the TSC is highly dependent on RSOC, magnetization strength and 
its orientation, and singlet-triplet mixing of the gap parameter. It is 
observed that NCSC with moderate RSOC shows maximum conductance for a 
partially opaque barrier in presence of low external magnetic field.  For a 
strongly opaque barrier and a nearly transparent barrier a moderate value and 
a low value of field respectively are found to be suitable. 
Moreover, NCSC with large singlet component is appeared to be useful. 
In addition, for NCSC with large RSOC and low magnetization strength, a giant 
STMR ($\%$) is observed. We have also seen that the spin current is strongly
magnetization orientation dependent. With the increase in bias voltage spin 
current increases in transverse direction, but the component along the direction 
of flow is almost independent.
\end{abstract}


\maketitle

\section{Introduction}
Spintronic devices, such as Spin Valves (SVs) or Magnetic Tunnelling Junctions 
(MTJs) have received a lot of attention over the years due to the significant 
progress in fabrication techniques. Traditional MTJs are composed of two 
ferromagnets in close proximity, normally separated by an insulator or a normal
metal. When a current is allowed to flow, it interact with the exchange field 
of the first ferromagnet and induces a polarization in the spin degrees of 
freedom. The second ferromagnet is introduced as spin detector, where spin 
current is measured \cite{johnson,jedema,wolf,casanova,chung}. The spin 
transport properties in these SVs are controlled mainly by the charge current, 
the relative orientation of the magnetization components and the external 
magnetic field. Moreover, depending upon the orientation of the magnetization 
i.e. parallel or anti parallel to the ferromagnetic regions, these hybrid
structures display Giant Magneto-Resistance (GMR) effect  
\cite{binasch,baibich,parkin,chappert} and hence have a great potential to be 
used as a non volatile magnetic memories, sensors for harddisk drives etc.  

Over the last two decades, the interplay between ferromagnetic and
superconducting order potentially enhances the interest in exploring 
Ferromagnet$|$Superconductor (F$|$S) hybrid structures for low temperature 
spintronic applications \cite{eschrig1,eschrig2,zutic,buzdin,blamire,
bergeret,kadigrobov,golubov,zhu11,petrashov,flokstra,slonczewski,berger,
halterman1,halterman2}. The discovery of phenomena like proximity effect \cite{zutic,buzdin,petrashov,kadigrobov}, long distance transport of magnetization \cite{zutic,buzdin,petrashov,kadigrobov}, spin injection \cite{flokstra},
Spin Transfer Torque (STT) \cite{slonczewski,berger}, and triplet correlation \cite{halterman1,halterman2} in F$|$S hybrid structures boosted up the 
superconducting spintronics research. Moreover, introduction of superconductor 
as a spacer in SVs provides the following advantages: (1) Superconducting 
spintronics devices can intricate strong proximity effect between the 
superconductor and the ferromagnet \cite{zutic,buzdin,petrashov,kadigrobov}, 
hence it gains a lot of interest from application point of view. 
(2) Superconducting SVs also reduce the consumption of energy and hence can 
be highly useful to fabricate ultra fast cryogenic magnetic memory devices 
\cite{golubov,zhu11}. (3) Furthermore, unconventional superconductors can also 
support polarized current \cite{bergeret,blamire,eschrig1,eschrig2} and hence 
it can be the potential candidate for superconducting spintronic devices.

Traditionally, conventional superconductors are introduced as a spacer in MTJs.
However, conventional superconductors are highly irreconcilable with 
ferromagnetism as the exchange coupling of a ferromagnet destroy the singlet 
pairing of the Cooper pairs \cite{bardeen}. On the other hand, 
superconductivity results
from the triplet pairing of the Cooper pairs can coexist with ferromagnetism 
and hence they are the prime candidates of F$|$S$|$F SVs. From the point of 
view of Cooper pairs, two symmetries play most pivotal role: the symmetry
of centre of inversion and the symmetry of time reversal. In absence of any of 
them the pairing can appear in an unconventional form. Over the last two 
decades, many heavy fermion compounds have been discovered which lack the 
center of inversion and hence they show unconventional superconductivity 
\cite{saxena, aoki, pfleiderer, huy, bauer1, bauer2, bauer3, motoyama, 
kawasaki, akazawa, anand1, smidman, anand2, yuan, togano, badica, matthias, 
singh, pecharsky, hillier, bonalde, yogi, ali, xu, flouquet, 
nandi,samokhin,sergienko,frigeri,fujimoto1,fujimoto2}.  Due to the lack of 
inversion centre, the Noncentrosymmetric Superconductors (NCSCs) are the 
candidates 
of prime concern since the last two decades.  Though NCSC had a great 
potential to be used in SVs from fundamental physics point of view, the field received a 
significant attention only after the discovery of unconventional 
superconductivity in CePt$_3$Si \cite{bauer1,bauer2,bauer3}. Soon many heavy 
fermion systems had been reported which show unconventional superconductivity 
due to the lacks center of inversion \cite{motoyama, kawasaki, akazawa, 
anand1, smidman, anand2, yuan, togano,badica, matthias, singh, pecharsky, 
hillier, bonalde, yogi, ali, xu,samokhin,sergienko,frigeri,fujimoto1,fujimoto2, 
flouquet, nandi}. A few of them are LaPt$_3$Si, La(Rh,Pt,Pd,Ir)Si$_3$, 
LaNiC$_2$, Li$_2$(Pt,Pd)$_3$B, UIr, Cd$_2$Re$_2$O$_7$, Re$_6$Zr, PbTaSe$_2$, 
etc. As the inversion symmetry is absent in NCSCs, hence parity is no longer 
remain conserved. Furthermore, due to the absence of inversion center in NCSC,
it offers strong Antisymmetric Spin-Orbit Coupling (ASOC). Inevitably, the 
Fermi surface split and the ground state of an NCSC exhibit a mixed pairing 
states consists of both spin singlet and spin triplet components. Thus the 
role of Spin-Orbit Coupling (SOC) is very significant in NCSCs. Since, SOC is 
antisymmetric and unconventional in NCSCs, hence it should be Rashba type SOC.  

Rashba Spin-Orbit Coupling (RSOC) \cite{bychkov,molenkamp,gorkov} is a 
symmetry dependent unconventional pairing 
$(\boldsymbol{\sigma}\times\boldsymbol{k})$ arises at the F$|$NCSC interface. 
Recent theoretical and experimental works indicate that RSOC is not only 
anticipated in the the mixing ratio of pairing states in NCSC but it also have 
the ability to tune spin triplet state from spin singlet state if Pd is 
replaced by Pt \cite{togano,yuan,badica} in the NCSC compound 
Li$_2$(Pd,Pt)$_3$B. This feature had also been extensively studied in many 
other NCSCs and the results significantly indicate that RSOC can have 
remarkable role in superconducting pairing mechanism. In general in most of 
cases, s-wave pairing dominates \cite{bauer4,wakui,ribeiro} in NCSCs, however 
it was also observed that for NCSC compound with comparatively low SOC, triplet
pairing state dominate over singlet pairing 
\cite{biswas,kuroiwa,chen1,chen2,wu}. Moreover, it was also found that Andreev 
reflection \cite{wu2,wu1,hashimoto,shigeta,beiranvand1,beiranvand2,halterman3,
romeo11} can also be tuned by RSOC and can be controlled by magnetization in 
graphene junction consists of ferromagnet and superconductor 
\cite{beiranvand1,beiranvand2,halterman3}. With the above mentioned 
motivations, it is necessary to investigate the role of RSOC in superconducting
pairing states in NCSCs and thereby in spin transport mechanism at F$|$NCSC 
junction of a F$|$NCSC$|$F SV.

To understand feasibility of ferromangetism and superconductivity, and also 
the role of magnetization in F$|$S hybrids, many superconducting 
spintronic SVs had been extensively studied theoretically
\cite{olthof,tagirov,zhu,fominov1,fominov2,acharjee,
fominov3,alidoust1,alidoust2,halterman4,romeo} and experimentally\cite{khaydukov,khaydukov1,leksin,nowak,gu,jara,zdravkov,srivastava,keizer,kuerten}.
Previous works \cite{olthof,tagirov,zhu,fominov1,
fominov2,fominov3,alidoust1,alidoust2,halterman4,romeo,khaydukov,khaydukov1,leksin,nowak,
 gu,jara,zdravkov,srivastava,acharjee,keizer,kuerten} on F$|$S heterostructure 
and superconducting SVs strongly indicates that the superconducting critical 
temperature is dependent on the orientation of magnetization. Recent 
experimental works predict that the formation of Andreev Bound State (ABS) 
\cite{keizer} and the flow of supercurrents \cite{kuerten} in SVs can be tuned 
and control via orientation of magnetization. Furthermore, it was also reported
that at the ballistic junction of N$|$F$|$Triplet SC in 
Sr$_2$RuO$_4$\cite{olthof}, the orientation of magnetization can even control 
the pairing potential.

Transport properties in several F$|$Singlet SC \cite{linder1,bozovic1,bozovic2},
F$|$Triplet SC \cite{cheng, cheng1,zutic1,zutic2,tanaka,moen1} 
and F$|$NCSC \cite{linder,iniotakis,kashiwaya,acharjee1}
heterostructures had been studied earlier. However, 
the role of magnetization in spin transport,
spin current \cite{moen2,an,sun,halterman5,linder5,halterman6}
and its interplay with pair potential, magnetic field and 
RSOC of NCSC in F$|$NCSC$|$F SV is still need to be understood.
Motivating from the previous results, in this work we study the spin transport
mechanism at F$|$NCSC junction of a F$|$NCSC$|$F SVs. More specifically, we 
have investigated the Tunnelling Spin Conductance (TSC), Spin Tunnelling 
Magneto-Resistance (STMR) and spin current for an F$|$NCSC$|$F SV architecture. 
To understand the interplay of RSOC with magnetization and the mixed pair 
potential of NCSC, we have studied TSC, STMR and spin current
considering each of those parameters extensively. Moreover, to make the setup 
experimentally reliable, the effect of an in-plane magnetic field and Fermi 
Wavevector Mismatch (FWM) are also investigated.

The paper is organized as follows. We briefly discuss theoretical framework 
of the proposed setup in the Sec. II. The results  and the discussions are 
presented in Sec. III. Finally, we summarize our work in Sec. IV.

\begin{figure}[hbt]
\centerline
\centerline{ 
\includegraphics[scale=0.5]{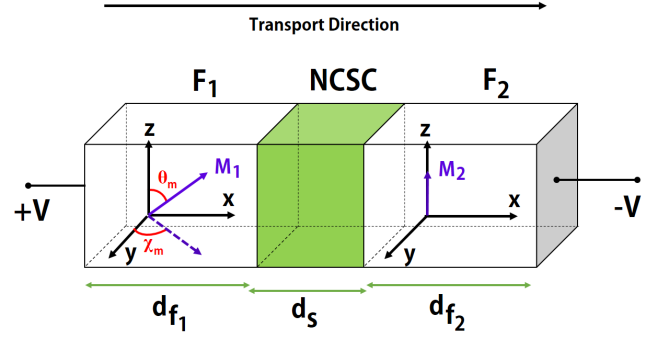}
\vspace{0.1cm}
}
\caption{ The schematic overview of the proposed experimental setup. In this 
F$|$NCSC$|$F spin valve architechture, the left F$_1$-layer represents a soft 
ferromagnet with exchange field $\boldsymbol{h} = h_0(\sin\theta_m \cos\chi_m, 
\sin\theta_m\sin\chi_m, \cos\theta_m)$, where $\theta_m$, $\chi_m$ 
respectively represent the polar and azimuthal angle of magnetization. The 
right F$_2$-layer represents a hard ferromagnet with a fixed orientation of 
magnetization. The middle NCSC-layer represents a noncentrosymmetric 
superconductor. In this work, x-direction is assumed to be as the direction of 
transport.}
\label{fig1}
\end{figure}

\section{Theory}
\subsection{Model and formalism}
We use the standard Bogoliubov-de Gennes (BdG) formalism to describe the 
behaviours of the Electron-Like-Quasi (ELQ) particle and
Hole-Like-Quasi (HLQ) particle amplitudes with spin $\sigma$.  
The schematic overview of the proposed setup is shown in Fig.\ref{fig1}.
For our setup, the BdG equations can be written as
 \begin{equation}
 \label{eq1}
 \left(
\begin{array}{cc}
 \hat{\mathcal{H}}_0  & \hat{\Delta }_{\alpha \beta }(r,r') \\
 \hat{\Delta }_{\alpha \beta }^{\dagger }(r,r') & -\hat{\mathcal{H}}_0^{\dagger } \\
\end{array}
\right)
\left(\begin{array}{c}
u_{n\sigma}  \\
v_{n\sigma} \\
\end{array}\right) = \epsilon_n
\left(\begin{array}{c}
u_{n\sigma}  \\
v_{n\sigma} \\
\end{array}\right)
 \end{equation}
where, $u_{n\sigma}$ and $v_{n\sigma}$ are the wavefunctions of ELQ particles 
and HLQ particles respectively. The hat sign represents a $2\times2$ matrix, 
$\epsilon_n$ are the energy eigenvalues can be obtained by diagonalizing the 
BdG Hamiltonian in the respective layers.  $\hat{\mathcal{H}}_0$ is 
the single particle Hamiltonian of the system, defined as
\begin{equation}
\label{eq2}
\hat{\mathcal{H}}_0 = (\mathcal{H'} + \mathcal{B})\hat{I} + 
\hat{\mathcal{H}}_R - \boldsymbol{h}\cdot\boldsymbol{\sigma}. 
\end{equation}
The first term in this  Eq.(\ref{eq2}) can be written as 
$\mathcal{H}' = -\frac{\nabla^2}{2} + E_{Fi} \,+\, U_0\,\delta(x)$,  where 
$-\frac{\nabla^2}{2}$ and $E_{Fi}$ are respectively the single particle 
kinetic energy and the Fermi energies in the respective layers. Here, we use 
standard units, viz., $\hbar = m = \mu = 1$. $U_0\delta(x)$ is the delta like 
barrier potential appears at F$|$NCSC interface with $U_0$ as the 
strength of spin independent barrier potential. $\mathcal{B}$ is the in-plane 
external magnetic field and $\hat{\mathcal{H}}_R$ is the RSOC, can be defined 
as \cite{wu,wu1,cheng1,acharjee1} 
\begin{equation}
\label{eq3}
\hat{\mathcal{H}}_R (\boldsymbol{k}) = \lbrace U_R\, \hat{e}_x\cdot
(\boldsymbol{\sigma}\times\boldsymbol{k})\rbrace
\end{equation}
where, $U_R$ represents the strength of RSOC, $\hat{e}_x$ is an unit vector 
directed normal to the interface and 
$\boldsymbol{k}$ = $-i\boldsymbol{\nabla}$. The last term in Eq.(\ref{eq2}) 
represents the exchange interaction, where $\boldsymbol{h}$ and 
$\boldsymbol{\sigma}$ are the exchange field and Pauli's spin matrices 
respectively.
 
The gap matrix $\hat{\Delta }_{\alpha \beta }(r,r')$ appear in the 
Eq.(\ref{eq1}) is the mixture of singlet (S) and triplet (T) components 
for a NCSC. It has the following form \cite{linder,acharjee1}:
\begin{equation}
\label{eq4}
 \hat{\Delta }_{\alpha \beta }(r,r') =
\left(
\begin{array}{cc}
 \Delta _{\uparrow\uparrow}(r,r') & \Delta _{\uparrow\downarrow}(r,r') \\
 \Delta _{\downarrow\uparrow}(r,r') & \Delta _{\downarrow\downarrow}(r,r') \\
\end{array}
\right)
\end{equation}

In general, $\Delta _{\uparrow\downarrow}(r,r')$  appearing in Eq.(\ref{eq4}) is a superposition
of the singlet (S) and the triplet (T) components that satisfies the following 
conditions:
\begin{align}
\Delta _{\uparrow\downarrow}(r,r')& = \Delta^S _{k\uparrow\downarrow}(r,r') +
 \Delta^T _{k\uparrow\downarrow}(r,r'),\\
\Delta^T _{k\uparrow\downarrow}(r,r')& = \Delta^T _{k\downarrow\uparrow}(r,r'),\\
\Delta^S _{k\uparrow\downarrow}(r,r')& = -\Delta^S _{k\downarrow\uparrow}(r,r').
\end{align}
 
Thus in view of Eqs.(\ref{eq1}), (\ref{eq2}), (\ref{eq3}) and (\ref{eq4}) the 
BdG equation in an extended form can be written as
 \begin{widetext}
 \begin{equation}
\label{eq8}
\left(
\begin{array}{cccc}
 -h_z+\mathcal{H}'+ \mathcal{B} & g_{k_-}-h_{\text{xy}} & \Delta^T _{k\uparrow
 \uparrow} & \Delta^S _{k\uparrow\downarrow}+\Delta^T _{k\uparrow\downarrow} \\
 g_{k_+}-h^*{}_{\text{xy}} & h_z+\mathcal{H}'+ \mathcal{B} & -\Delta^S _{k\uparrow
 \downarrow}+\Delta^T _{k\uparrow\downarrow} & \Delta^T _{k\downarrow\downarrow} \\
 {\Delta^T_{k\uparrow\uparrow}}^\dagger &  -{\Delta^S _{k\uparrow\downarrow}}^\dagger 
 + {\Delta^T _{k\uparrow\downarrow}}^\dagger & h_z-\mathcal{H}'- \mathcal{B} & g_{k_+}
 -h^*{}_{\text{xy}} \\
 {\Delta^S _{k\uparrow\downarrow}}^\dagger+{\Delta^T _{k\uparrow\downarrow}}^\dagger &  
 {\Delta^T_{k\downarrow\downarrow}}^\dagger & g_{k_-}-h_{\text{xy}} & -h_z-\mathcal{H}'- \mathcal{B} \\
\end{array}
\right)
\left(\begin{array}{c}
u_{n\uparrow}  \\
u_{n\downarrow}  \\
v_{n\uparrow} \\
v_{n\downarrow} \\
\end{array}\right) = \epsilon_n
\left(\begin{array}{c}
u_{n\uparrow}  \\
u_{n\downarrow}  \\
v_{n\uparrow} \\
v_{n\downarrow} \\
\end{array}\right)
  \end{equation}
\end{widetext}
where, $h_{xy} = h_0(\sin\theta_m \cos\chi_m - i \sin\theta_m\sin\chi_m)$, 
$h_z = h_0 \cos\theta_m$ and $g_{k\pm} = U_R(k_x \mp i k_y)\Theta(x)$. Here
$\Theta(x)$ is the Heavyside step function can be defined as 
\begin{equation}
\label{eq9}
\Theta(x) = \left\{\begin{array}{rc} 
0, & x <0,\\
1, & x \ge 0.
\end{array} \right.
\end{equation} 

In order to obtain the momenta $k^+ (k^-)$ for the electron(holes) in the 
soft F-layer, we diagonalize the BdG Hamiltonian appearing in the 
Eq.(\ref{eq8}). On diagonalizing, which can be found as  
\begin{equation}
\label{eq10}
k^{\sigma} =  k_{FF}\sqrt{1 + Z_0 -\sigma Z_R \sin\theta_F -\sigma X + B \pm Z_1} 
\end{equation}
where, $\sigma = \pm 1$ represent two different orientations of the 
spin and $k_{FF}$ represent the Fermi momentum of electron and holes at 
F-layer. Here, we define $Z_0 = \frac{2U_0}{k_{FF}}$, $Z_R = 2U_R$, 
$X = \frac{M_1}{E_{FF}}$, $B =\frac{\mathcal{B}}{E_{FF}}$ and 
$Z_1 = \frac{E}{E_{FF}}$ with $M_1$ as the strength of magnetization of the 
left ferromagnetic layer, $X$ as the strength of magnetization per
unit Fermi energy and $E_{FF}$ as the Fermi energy of electrons and holes in 
that region.

In a similar way, we represent the momenta of the ELQ(HLQ) particles in the 
superconducting region by $q^+$($q^-$), which are defined as
\begin{equation}
\label{eq11}
q^\pm = \sqrt{2(E_{FS}- B \pm\sqrt{E^2-\Delta_{\alpha\beta}^2})}
\end{equation}
where, $E_{FS}$ is the Fermi energy of ELQ and HLQ particles in the 
superconducting region. Since, in the tunnelling mechanism the parallel 
component of momenta is conserved. So we can write,
\begin{equation}
\label{eq12}
k^+\sin\theta_{\text{F}} = k^-\sin\theta_{\text{A}} = q^+\sin{\theta_e} = q^-\sin\theta_h
\end{equation}
where, $\theta_F$ and $\theta_A$ are respectively are the angle of incidence 
of the electron in F-region and the Andreev reflected angle for the hole in 
the superconducting region. $\theta_e$ is the angle of refraction
for the ELQ particles, while $\theta_h$ is the angle of refraction for the 
HLQ particles.
 
Since, the Fermi energy in NCSC is quite different from a ferromagnet, so 
to characterize this, we introduce a FWM parameter $\lambda$.
Physically, it is a dimensionless parameter defined as the ratio 
of the Fermi momentum ($q_{FS}$)  in the superconducting region to Fermi
momentum ($k_{FF}$) in the Ferromagnetic region, i.e. 
$\lambda = \frac{q_{FS}}{k_{FF}}$. The wave function $\Psi_{\text{F}}(x)$ in 
F-layer with any arbitrary orientation of magnetization is given by
\begin{multline}
\label{eq13}
 \Psi_{\text{FM}}(x<0) = s_{\uparrow }[\delta_1 \hat{e}_1 + \delta_2 \hat{e}_2]e^{i k^{+}\cos\theta_Fx}
\\+s_{\downarrow }[-\delta_2^* \hat{e}_1 + \delta_1 \hat{e}_2]e^{ik^{-}\cos\theta_Fx}
\\+r_e^{\uparrow }[\delta_1 \hat{e}_1 + \delta_2 \hat{e}_2]e^{-i k^{+}S_1x}
\\+r_e^{\downarrow }[-\delta_2^* \hat{e}_1 + \delta_1 \hat{e}_2]e^{-i k^{+}S_2x}
\\+r_h^{\uparrow }[\delta_1 \hat{e}_3 + \delta_2 \hat{e}_4]e^{ik^{+}S_1x}
\\+r_h^{\downarrow }[-\delta_2^* \hat{e}_3 + \delta_1 \hat{e}_4]e^{ik^{-}S_2x}
\end{multline}
where, we define $\hat{e}_1 = (1,0,0,0)^T$, $\hat{e}_2 = (0,1,0,0)^T$, $\hat{e}_3 = (0,0,1,0)^T$, $\hat{e}_4 = (0,0,0,1)^T$, $\delta_1 = \cos\theta_m$, $\delta_2 = \sin\theta_m  e^{-i \chi_m}$, $S_1 = s_\uparrow \cos\theta_{\text{F}} +
s_\downarrow \cos\theta_{\text{A}}$ and $S_2 = s_\uparrow \cos\theta_{\text{A}} + s_\downarrow \cos\theta_{\text{F}}$. For up spin incident particle we choose 
$s_\uparrow $ = 1, $s_\downarrow $ = 0, while for a down spin particle 
$s_\uparrow $ = 0, $s_\downarrow $= 1. $\theta_m$ and $\chi_m$ are the polar 
angle and azimuthal angle of magnetization corresponding to the magnetization 
vector in the soft ferromagnetic layer. $r_e^{\uparrow }$ ($r_e^{\downarrow }$)
appearing in the Eq.(\ref{eq13}) are the normal reflection coefficients 
for upspin (downspin) electrons, while $r_h^{\uparrow }$ ($r_h^{\downarrow }$) 
are the Andreev reflection coefficients for the upspin (downspin) holes.

In a similar way, the wave function $\Psi_{\text{NCSC}}(x)$ in NCSC-layer can 
be written \cite{linder} as
\begin{multline}
\label{eq14}
\Psi _{\text{NCSC}}(x \geq 0)=
\frac{t_e{}^{\uparrow }}{\sqrt{2}}[u_+\Gamma_1 + v_+\Gamma_4] e^{iq_e^{+}\cos \theta _ex}
\\+ \frac{t_e{}^{\downarrow }}{\sqrt{2}}[u_-\Gamma_2 + v_-\Gamma_3] e^{iq_e^{-}\cos \theta _ex}
\\+ \frac{t_h{}^{\uparrow }}{\sqrt{2}}[v_+\Gamma_1 + u_+ \Gamma_4]e^{iq_h^{+}\cos \theta _hx} 
\\+ \frac{t_h{}^{\downarrow }}{\sqrt{2}}[v_-\Gamma_2 + u_- \Gamma_3]e^{i\text q_h^{-}\cos \theta _hx}
\end{multline}
where, $\Gamma_1 = (\hat{e}_1 + \hat{e}_2 e^{-i\phi})$,  $\Gamma_2 = (\hat{e}_1 - \hat{e}_2 e^{-i\phi})$, $\Gamma_3 = (\hat{e}_4 + \hat{e}_3 e^{-i\phi})$ and  
$\Gamma_4 = (\hat{e}_4 - \hat{e}_3 e^{-i\phi})$. $\phi$ is the superconducting 
phase factor, $ t_e^{\uparrow }$( $t_e^{\downarrow }$) corresponds to the 
transmission coefficients for up(down) spin of ELQ particles, while
$t_h^{\uparrow }$( $t_h^{\downarrow }$) represents the transmission 
coefficients for up(down) spin of HLQ particles. The amplitudes of 
wavefunctions of ELQ particles and HLQ particles are given by
\begin{equation}
\label{eq15}
u_\pm ( v_\pm) =  \frac{1}{\sqrt{2}}\sqrt{1+(-)\sqrt{1-\frac{|\Delta_s\pm\frac{\Delta_t}{2}|^2}{E^2}}}.
\end{equation}

The reflection coefficients ($r_e^\sigma $, $r_h^\sigma$) and the transmission 
coefficients ($t_e^\sigma $, $t_h^\sigma$) in the wavefunctions can be 
determined under the following boundary conditions:
\begin{eqnarray}
\label{eq16}
\Psi_\text{FM}(x = 0^-) = \Psi_\text{NCSC}(x = 0^+),\\
\label{eq17}
\nonumber
\partial_x \lbrace\Psi_\text{NCSC}(x = 0^+)-
\Psi_\text{FM}(x = 0^-)\rbrace =\\ 2U_{int}\Psi_\text{FM}(x=0)
\end{eqnarray}
where, $U_{int} = U_0\delta(x) + \mathcal{H}_R$ is the interacting potential.

\subsection{Calculation of Spin Conductance at the F$|$NCSC interface}
To calculate the spin conductance $G_S(E)$, we have used the extended  
Blonder - Tinkham - Klapwijk (BTK) approach \cite{blonder}. According to BTK 
formalism, the spin conductance $G_S^\uparrow(E,\theta_F)$ for an upspin 
incoming electron incident at an angle $\theta_F$ is
\begin{equation}
\label{eq18}
G_S^\uparrow(E,\theta_F) = \frac{1 + X}{2}(1 - |r_e^\uparrow|^2 + |r_e^\downarrow|^2 
+ |r_h^\uparrow|^2 - |r_h^\downarrow|^2),
\end{equation}
while for a downspin incoming electron, the the spin conductance $G_S^\downarrow(E,\theta_F)$ is
\begin{equation}
\label{eq19}
G_S^\downarrow(E,\theta_F) = \frac{1 - X}{2}(1 + |r_e^\uparrow|^2 - |r_e^\downarrow|^2 
+ |r_h^\uparrow|^2 - |r_h^\downarrow|^2).
\end{equation}
Thus, in view of this the angularly averaged spin conductance can be 
written as \cite{acharjee1,cheng1,blonder,linder,linder1}
\begin{equation}
\label{eq20}
G_S(E) = G_N^{-1} \int_{-\frac{\pi}{2}}^{\frac{\pi}{2}}  d\theta_F \cos\theta_F 
\lbrace G_S^\uparrow + G_S^\downarrow \rbrace,
\end{equation}
where $G_N$ is the tunnelling conductance for N$|$N (N for normal mattel) 
junction and has the following form:
\begin{equation}
\label{eq21}
G_N = \int_{-\frac{\pi}{2}}^{\frac{\pi}{2}} d\theta_F  \frac{4\cos^3\theta_F}{4\cos^2\theta_F + Z_0^2}.
\end{equation}

\subsection{Spin Tunnelling Magneto-Resistance (STMR)}
It is seen from Eqs.(\ref{eq18}) and (\ref{eq19}) that the spin 
conductance for spin-up particles is quite different from that of spin-down 
particles. Thus it generates a STMR. Moreover, it also seen from 
Eqs.(\ref{eq18}) and (\ref{eq19}) that the value STMR is depended on the 
magnetization strength (X). So in this work we have studied the STMR for 
different X and magnetic field (B). It is to be noted that STMR can be 
calculated by knowing the reflection and transmission coefficients from 
Eqs.(\ref{eq16}) and (\ref{eq17}) at the different spin subbands and 
then inserting them in spin conductance Eqs.(\ref{eq18}) and (\ref{eq19}). 
The STMR can be defined \cite{cheng1} as  
 \begin{equation}
 \label{eq22}
\text{STMR}  = \frac{G_S^{P}(E) - G_S^{AP}(E)}{G_S^{P}(E)}
 \end{equation}
where, $G_S^{P}(E)$ and $G_S^{AP}(E)$ respectively corresponds to the spin 
conductances at parallel and anti-parallel orientations.
 
\subsection{Spin Current ($\boldsymbol{S}$)} 
Due to the non-collinear orientation of magnetization in the two 
ferromagnetic layers of the F$|$NCSC$|$F SV, a spin current 
$\boldsymbol{S}$ is generated and flows through the system even in 
absence of a charge current. Thus  the
spin current can be totally controlled by the strength and the 
orientation of exchange fields. Moreover,  the 
spin currents in the ferromagnetic layers generate a torque which 
tends to rotate the magnetizations. The spin continuity 
equation can be written as
\begin{equation}
\label{eq23}
\partial_t\langle \boldsymbol{\eta}(x)\rangle 
+ \partial_x \boldsymbol{S}(x) = \boldsymbol{\tau}(x)
\end{equation}
where, $\partial_t = \frac{\partial}{\partial t}$, 
$\partial_x = \frac{\partial}{\partial x}$, 
$\boldsymbol{\tau}(x) = -2\langle \psi^\dagger(x) 
\big(\boldsymbol{\sigma} \times \boldsymbol{h}\big)\psi(x)\rangle$
is the Spin Transfer Torque (STT) and $\boldsymbol{\eta}(x)$ 
is the spin density operator related to magnetization as
$\boldsymbol{m}(x) = -\mu_B\langle\boldsymbol{\eta}(x)\rangle$. 
The spin density is in general has a tensor form since it 
has both direction of flow in real space and a direction in 
spin space. However, it can be reduced to vector form by 
the quasi-one-dimensional nature of the geometry. The spin 
current can be defined as 
\begin{equation}
\label{eq24}
\boldsymbol{S}(x) = -\frac{i}{2m}\langle\psi^\dagger(x)\boldsymbol{\sigma}
 \partial_x\psi(x) - \partial_x\psi^\dagger(x)\boldsymbol{\sigma} \psi(x)\rangle.
\end{equation}
we can write $\boldsymbol{S}$ in terms of quasi-particle amplitudes and 
energies using Bogoliubov transformations:
\begin{equation}
\label{eq25}
\psi_\sigma(r) = \sum_n[u_{n\sigma}(r)\gamma_n+\eta 
v^{\ast}_{n\sigma}(r)\gamma^{\dagger}_n],
\end{equation} 
where $u_{n\sigma}$ and $v_{n\sigma}$ are the quasi-particle and quasi-hole 
amplitudes. $\gamma_n$ and $\gamma^{\dagger}_n$ are Bogoliubov quasi-particle 
annihilation and creation operators respectively, which satisfy the following 
expectation values: 
$\langle\gamma^\dagger_n\gamma_m\rangle = \delta_{mn}f_n$,
$\langle\gamma_m\gamma^\dagger_n\rangle = \delta_{nm}(1-f_n)$ 
and $\langle \gamma_n\gamma_m\rangle = 0$. Here,
$f_n = \big[ \exp\big(\frac{\epsilon_n}{2T}\big)+1 \big] ^{-1}$ is 
the Fermi function which is dependent on temperature T and 
quasi-particle energy $\epsilon_n$. Calculating the reflection, transmission 
coefficients and inserting Bogoliubov 
transformations (\ref{eq25}), the components of spin current (\ref{eq24}) 
\cite{halterman5,moen2,linder5} can be represented in terms of quasi-particle 
amplitude as
\begin{multline}
\label{eq26}
S_x(x) = -\frac{i}{2m}\sum_n\Big[ f_n\Big\{u^{\ast}_{n\uparrow}\partial_x u_{n\downarrow}
 + u^{\ast}_{n\downarrow} \partial_x u_{n\uparrow} \\ - 
u_{n\downarrow}\partial_x u^{\ast}_{n\uparrow} -  
u_{n\uparrow}\partial_x u^{\ast}_{n\downarrow}\Big\}
- (1-f_n)\Big\{v_{n\uparrow}\partial_x v^{\ast}_{n\downarrow}
\\+v_{n\downarrow}\partial_x v^{\ast}_{n\uparrow}
-v^{\ast}_{n\uparrow}\partial_x v_{n\downarrow}
-v^{\ast}_{n\downarrow} \partial_x v_{n\uparrow}\Big\}\Big],
\end{multline}

\begin{multline}
\label{eq27}
S_y(x) = -\frac{1}{2m}\sum_n\Big[ f_n\Big\{u^{\ast}_{n\uparrow}\partial_x 
u_{n\downarrow} - u^{\ast}_{n\downarrow} \partial_x u_{n\uparrow} \\ - 
u_{n\downarrow}\partial_x u^{\ast}_{n\uparrow} +  
u_{n\uparrow}\partial_x u^{\ast}_{n\downarrow}\Big\}
- (1-f_n)\Big\{v_{n\uparrow}\partial_x v^{\ast}_{n\downarrow}
\\-v_{n\downarrow}\partial_x v^{\ast}_{n\uparrow}
+v^{\ast}_{n\uparrow}\partial_x v_{n\downarrow}
-v^{\ast}_{n\downarrow} \partial_x v_{n\uparrow}\Big\}\Big],
\end{multline}

\begin{multline}
\label{eq28}
S_z(x) = -\frac{i}{2m}\sum_n\Big[ f_n\Big\{u^{\ast}_{n\uparrow}\partial_x 
u_{n\uparrow} - u_{n\uparrow} \partial_x u^{\ast}_{n\uparrow} \\ - 
u^{\ast}_{n\downarrow}\partial_x u^{\ast}_{n\downarrow} +  
u_{n\downarrow}\partial_x u_{n\downarrow}\Big\}
- (1-f_n)\Big\{-v_{n\uparrow}\partial_x v^{\ast}_{n\uparrow}
\\+v^{\ast}_{n\uparrow}\partial_x v_{n\uparrow}
+v_{n\downarrow}\partial_x v^{\ast}_{n\downarrow}
-v^{\ast}_{n\downarrow} \partial_x v_{n\downarrow}\Big\}\Big].
\end{multline}

\section{Results and Discussion}
\subsection{Spin Conductance Spectra at the F$|$NCSC interface}
In this work, we study the spin transport quantities, more specifically
TSC ($G_S$), STMR and the spin current ($\boldsymbol{S}$) for a F$|$NCSC$|$F 
SV. We have 
plotted the spin conductance $G_S (E)$ from the equation (\ref{eq20}) as a 
function of baising energy $E$ scaled by the gap amplitude $\Delta$ of NCSC. 
Since NCSCs posses a mixed pairing state, so to understand the interplay of 
pairing symmetry on spin transport we have introduced the gap amplitude 
parameter $|\Delta_\pm|$, where $|\Delta_\pm| = 
|\Delta_s\pm\frac{\Delta_t}{2}|$. For most of our analysis we choose 
$\Delta_s =\frac{\Delta_t}{3}$. However to understand the impact of 
singlet-triplet mixing ratio on the spin conductance, we have also considered 
different mixing ratios too. Furthermore, to study the spin conductance spectra
we set the magnetization strength, polar angle and the azimuthal angle of 
magnetization respectively as $X = 0.9$, $\theta_m = 0.25\pi$ and 
$\chi_m = 0.5\pi$. It is to be noted here that the densities of the local 
charge carries in different regions of F$|$NCSC heterostructure are different. 
Again, the Fermi momenta for a ferromagnet is also quite different from a NCSC. So to incorporate this point we introduce a dimensionless parameter 
$(\lambda)$, which characterizes the FWM at the different regions. 
Though $\lambda$ can have any arbitrary values, however for high temperature
superconductors FWM is found to be less than unity \cite{linder1}. Hence
for our calculation of spin conductance \cite{acharjee1,linder,linder1}, STMR 
and spin current, we set $\lambda = 0.5$. 

In most of the earlier works on tunnelling spectroscopy
\cite{linder1,bozovic1,bozovic2,cheng, cheng1,zutic1,zutic2,
tanaka,moen1,linder,iniotakis,kashiwaya,acharjee1}, it was found that the 
barrier thickness play a very significant role on the transport mechanism. 
However, the role of barrier thickness on spin transport is yet to be 
understood. So, we study the variation of spin conductance with applied 
baising energy for different RSOC ($Z_R$), magnetization strength $(X)$ and 
in-plane magnetic field ($B$) considering initially a partially opaque 
barrier with $Z_0 = 1.0$ and a highly opaque barrier of barrier thickness  
$Z_0 = 2.0$. The result of which are presented in Figs.\ref{fig2} and 
\ref{fig3} respectively. Furthermore, for a complete understanding of the 
quantum tunnelling mechanism and the role of barrier thickness, we have also 
considered a highly transparent barrier with barrier width $Z_0 = 0$ in 
Fig.\ref{fig4}. 

\subsubsection*{Effect of Rashba Spin Orbit Coupling (RSOC)}
\begin{figure}[hbt]
\centerline
\centerline{
\includegraphics[scale = 0.42]{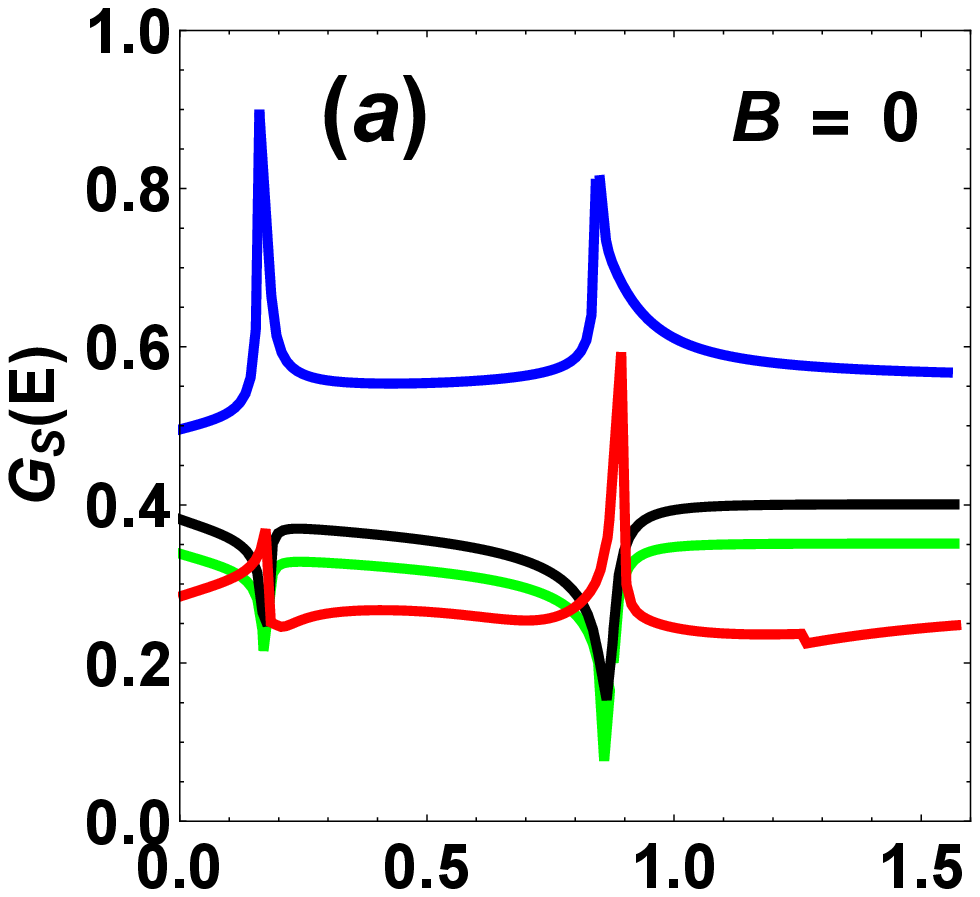}
\hspace{0.15cm}
\vspace{0.25cm}
\includegraphics[scale = 0.42]{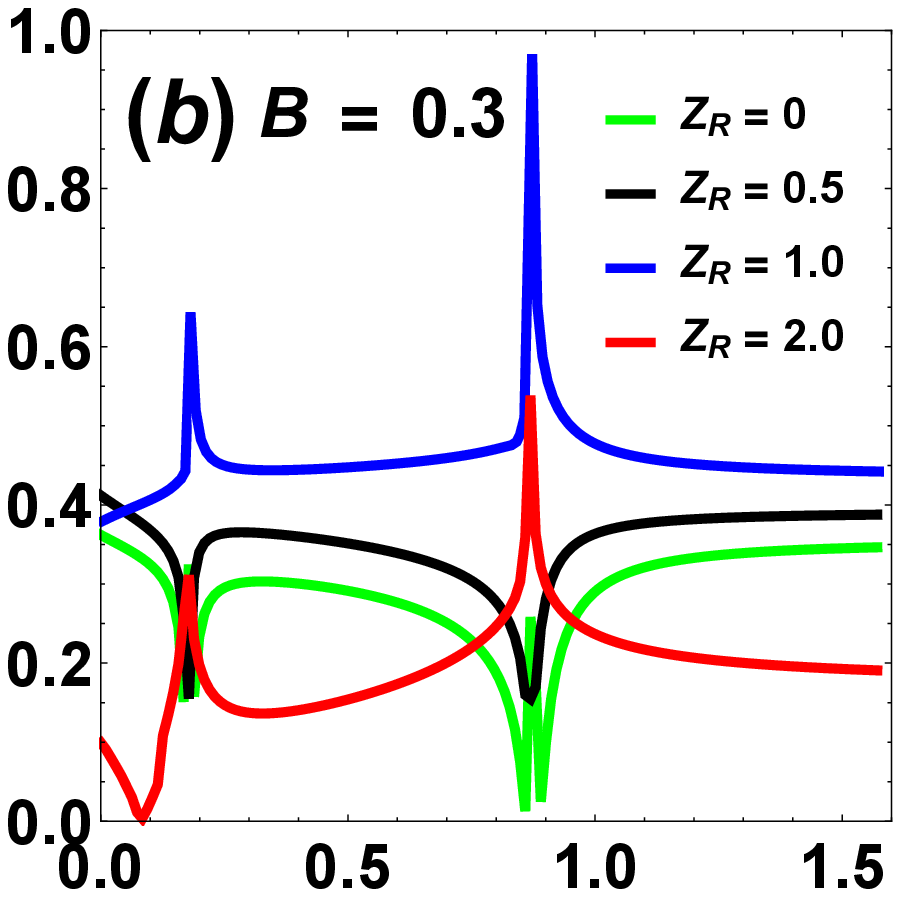}
\vspace{0.25cm}
\hspace{0.15cm}
\includegraphics[scale = 0.42]{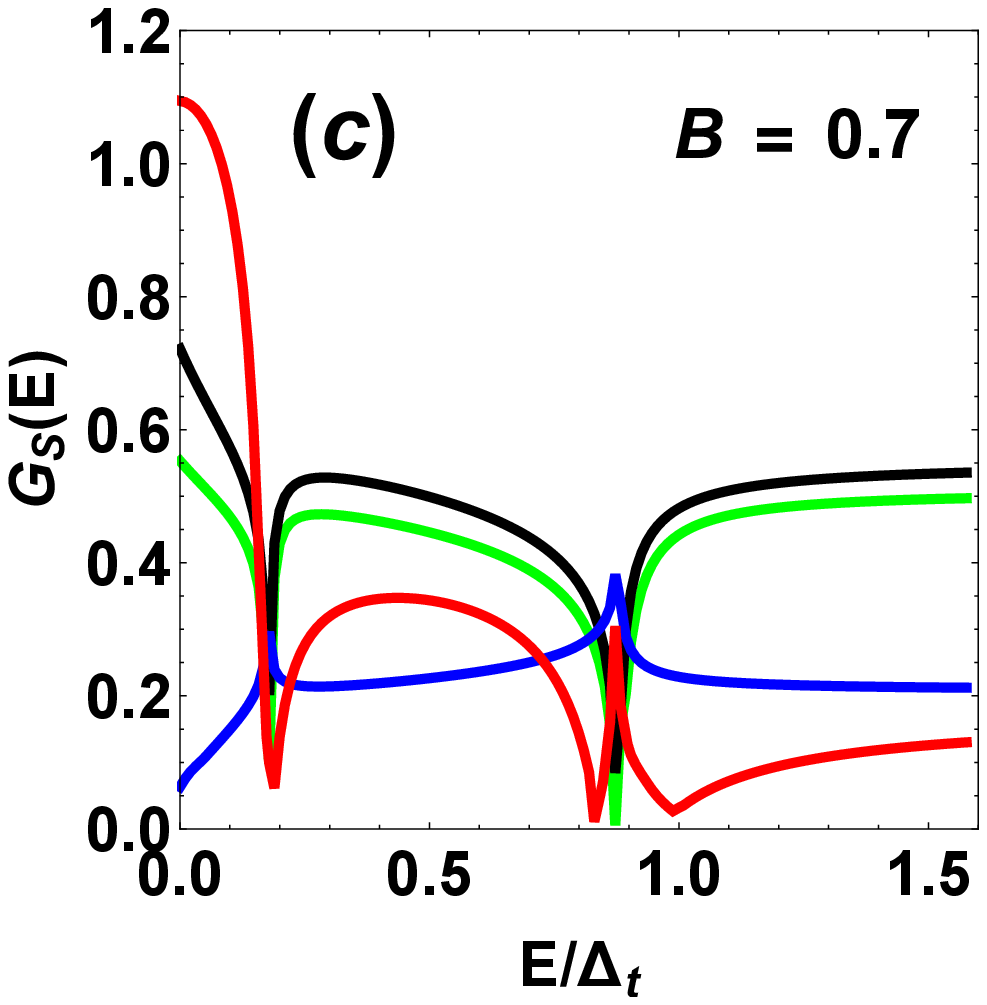}
\hspace{0.3cm}
\includegraphics[scale = 0.42]{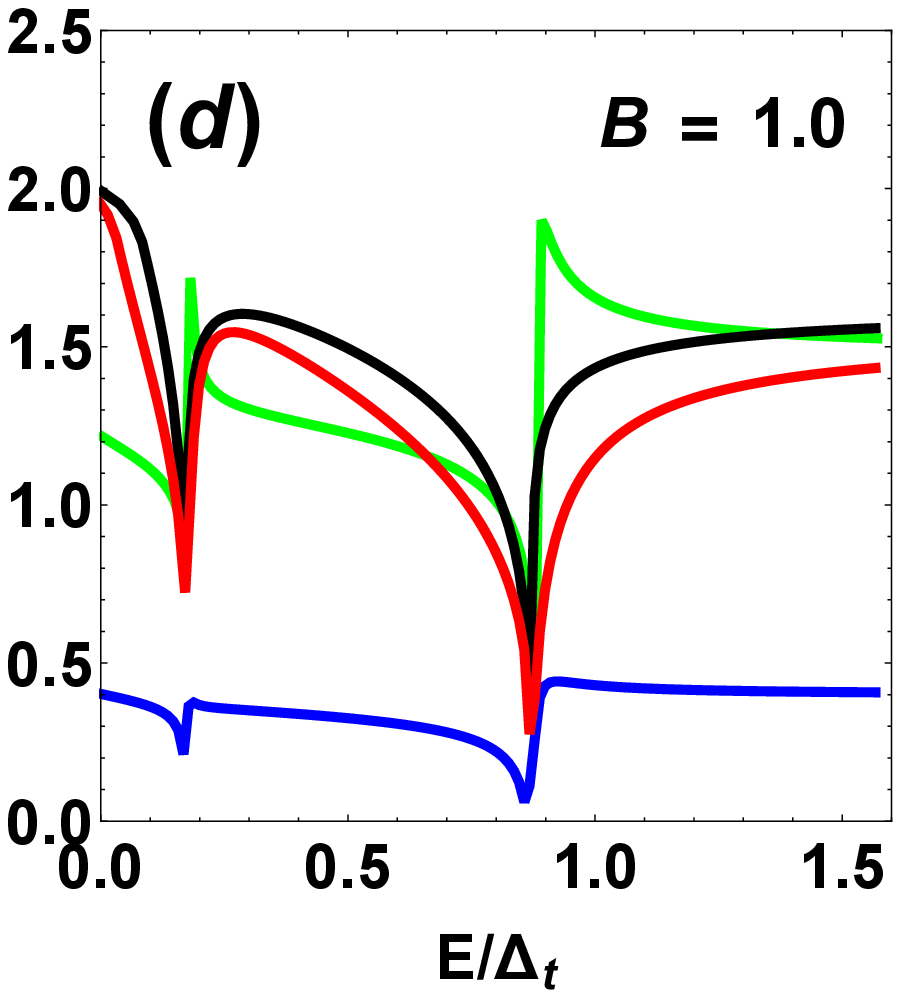}
}
\caption{Spin conductance spectra for different values of $B$ and for $X=0.9$ 
considering $\Delta_s =\frac{\Delta_t}{3}$, $Z_0 = 1.0$, $\lambda = 0.5$, $\theta_m = 0.25\pi$ and $\chi_m = 0.5\pi$.}
\label{fig2}
\end{figure}

From the previous works, it is found that the role of RSOC in NCSC is too 
notable. Though RSOC has a very pivotal role in superconducting pairing, 
but its interplay with magnetization, magnetic field is still need to be 
understood. So in view of this in Fig.\ref{fig2} we study the effect of 
RSOC $(Z_R)$ on the spin conductance $G_S(E)$. For all our analysis on spin 
conductance we consider four different choices of RSOC viz., $Z_R = 0, 0.5, 
1.0$ and $2.0$. Moreover, to understand the significance of external in-plane 
magnetic field on spin conductance we have also considered four different 
in-plane magnetic field strengths viz., $B = 0, 0.3, 0.7$ and $1.0$ 
respectively in Figs.\ref{fig2}(a), \ref{fig2}(b), \ref{fig2}(c) and 
\ref{fig2}(d). 
It is seen that 
due to the formation of Andreev Bound States (ABS) nearer to the baising 
energies, $E = \Delta_-$ = $|\Delta_s - \frac{\Delta_t}{2}|$ 
and at  $\Delta_+$ = $|\Delta_s + \frac{\Delta_t}{2}|$, two sharp peaks in the 
spin conductance are observed for $Z_R = 1.0$ and $2.0$ in absence and for low 
magnetic field $B = 0.3$. It is to be noted that $Z_R = 1.0$ shows maximum
conductance in both the cases. The sharpness of the peak increases nearly at 
$E \sim \Delta_+ = 0.83 \Delta_t$, while it decreases at 
$E \sim \Delta_- = 0.17 \Delta_t$ for a very low value of $B = 0.3$. The 
situation is opposite at these two points in absence of $B$. Also the
sharpness is highest nealy at $E \sim \Delta_+ = 0.83 \Delta_t$ for $B = 0.3$. 
These results strongly indicate that in-plane magnetic field must have 
some significant role on the spin transport. With the further rise of $B$ to 
$0.7$ the sharpness of the peaks get decreased and finally for $B = 1.0$ two 
dips are seen for all choices of $Z_R$ as clear from Fig.\ref{fig2}(d).  
This is because in presence of strong in-plane magnetic field $B$ the exchange 
field becomes very strong, hence it becomes unfavourable for the 
superconducting pairing. Thus ABS get suppressed and is characterised by the 
dips in Fig.\ref{fig2}. It is also to be noted that for Rashba free case 
$Z_R = 0$ and for low RSOC value $Z_R = 0.5$, two sharp dips are also 
observed for all choices of magnetic field. However, for $Z_R = 0$, a sharp 
rise is seen in case of $B = 1.0$. Moreover, it is also observed from 
Fig.\ref{fig2}(d) that for $B = 1.0$ the spin conductance becomes maximum for 
$Z_R = 0.5$ and it sharply decreases as the biasing energy 
approaches the gap energies $\Delta_\pm$ for all choices of $Z_R$.

\begin{figure}[hbt]
\centerline
\centerline{
\includegraphics[scale = 0.42]{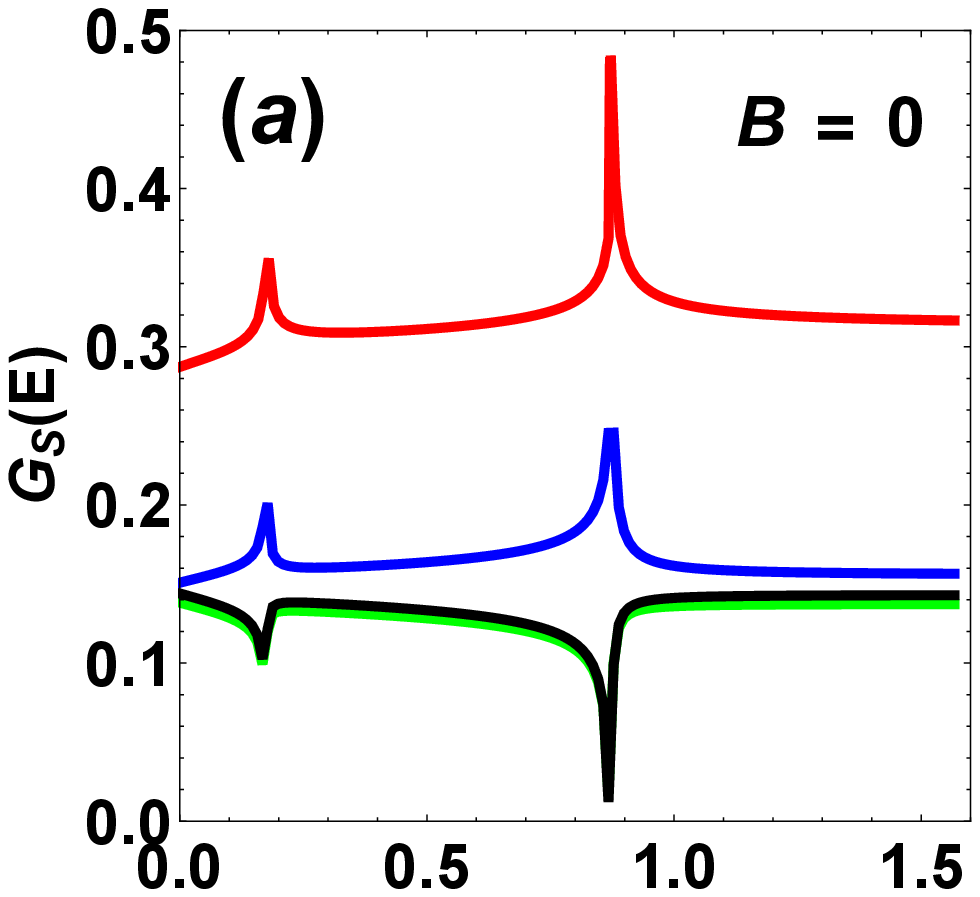}
\hspace{0.15cm}
\vspace{0.25cm}
\includegraphics[scale = 0.42]{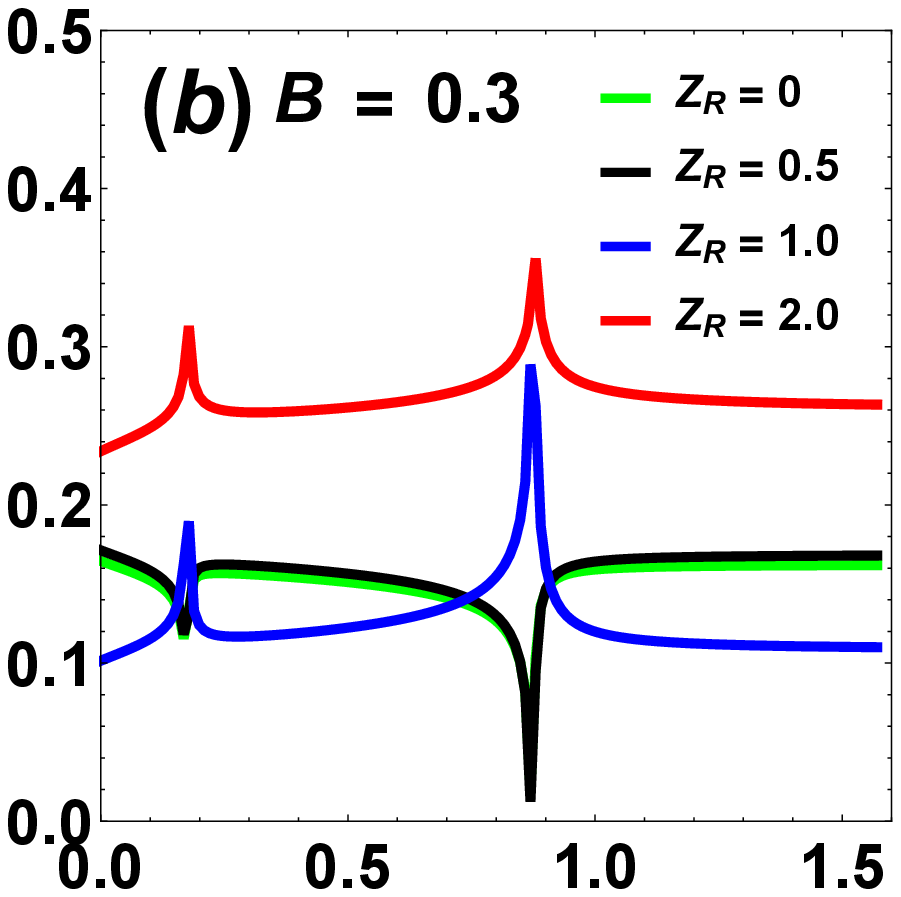}
\hspace{0.35cm}
\vspace{0.25cm}
\includegraphics[scale = 0.42]{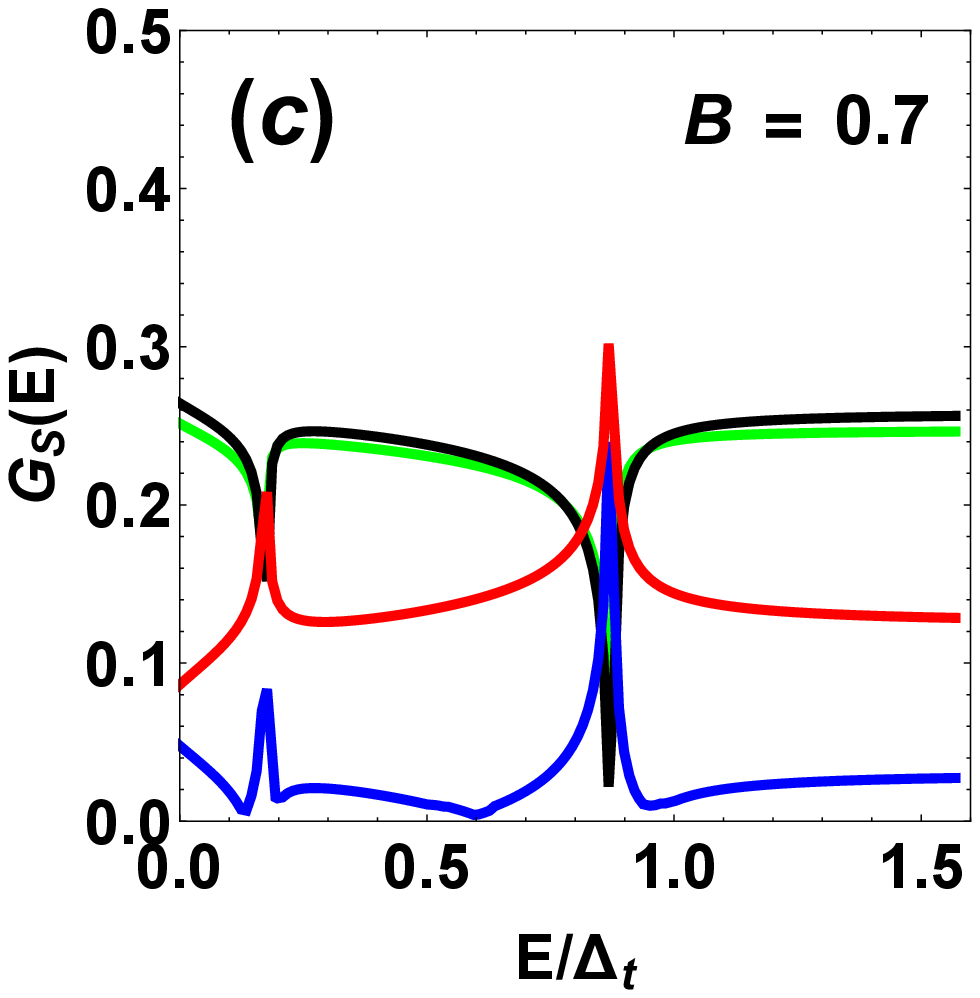}
\hspace{0.3cm}
\includegraphics[scale = 0.42]{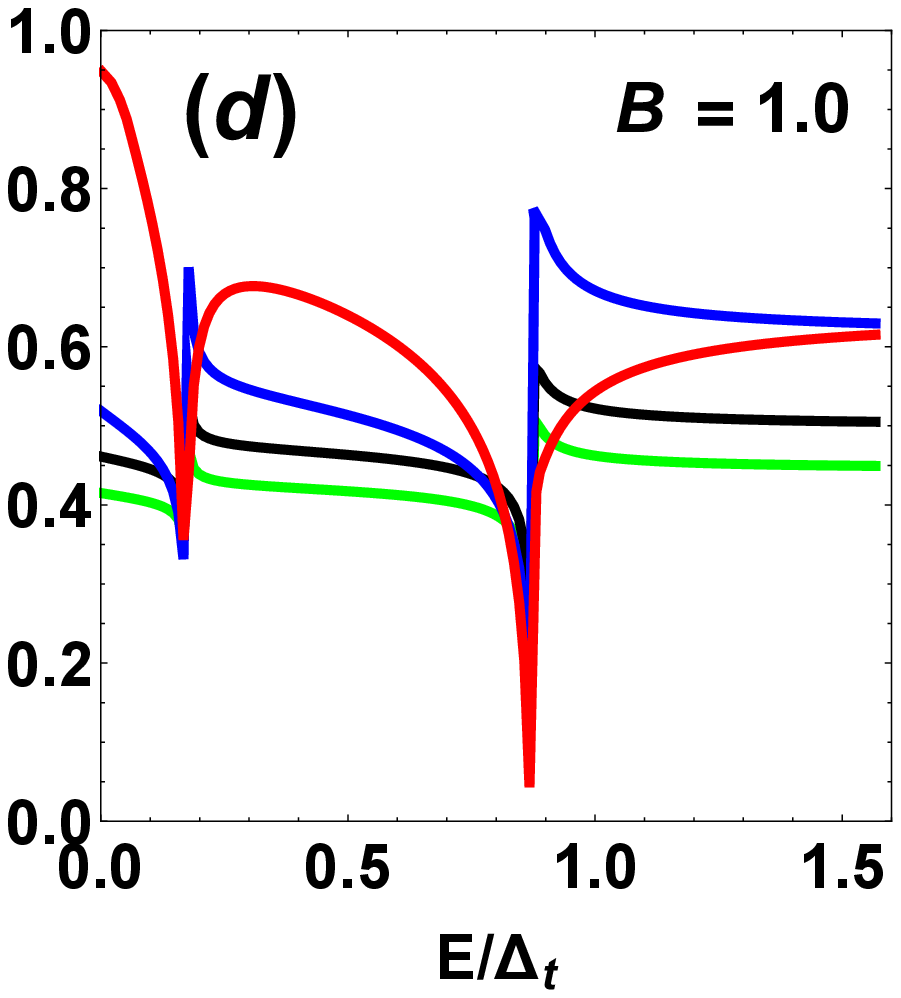}
}
\caption{Spin conductance spectra for different values of $B$ and for $X=0.9$ 
considering $\Delta_s =\frac{\Delta_t}{3}$, $Z_0 = 2.0$, $\lambda = 0.5$, 
$\theta_m = 0.25\pi$ and $\chi_m = 0.5\pi$.}
\label{fig3}
\end{figure}

\begin{figure}[hbt]
\centerline
\centerline{
\includegraphics[scale = 0.42]{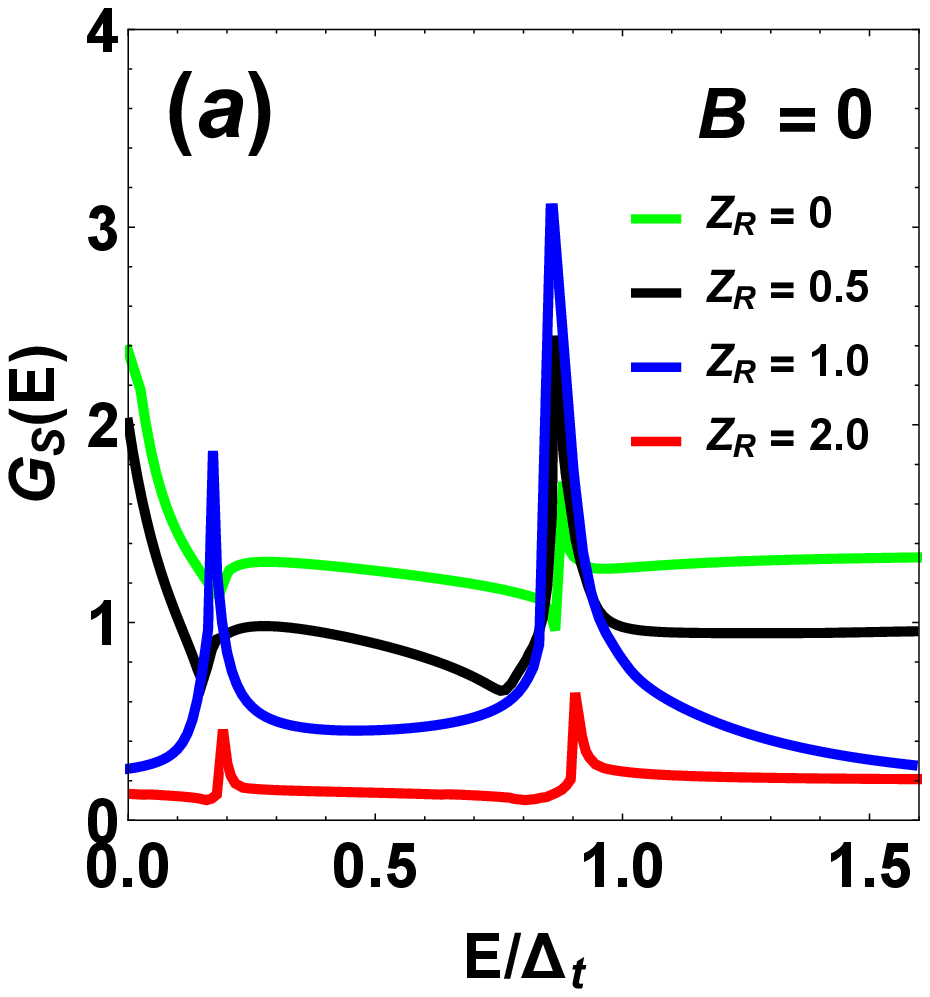}
\hspace{0.35cm}
\includegraphics[scale = 0.42]{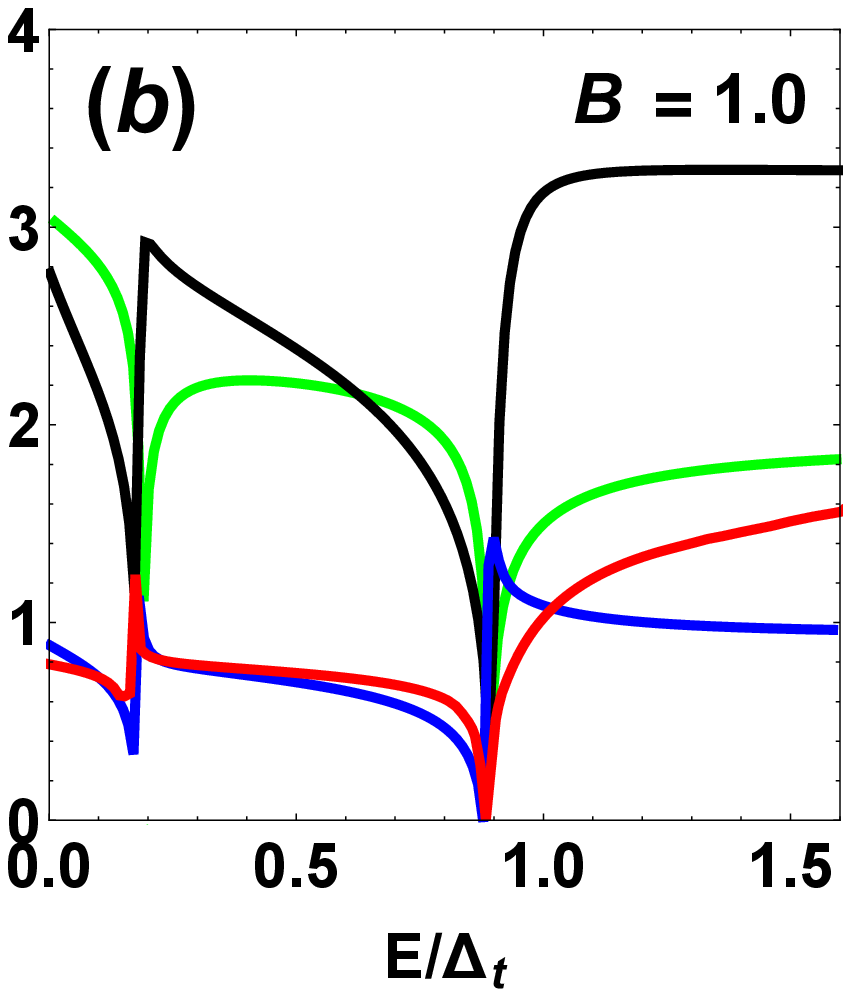}
}
\caption{Spin conductance spectra for different values of $B$ and for $X=0.9$
considering $\Delta_s =\frac{\Delta_t}{3}$, $Z_0 = 0$, $\lambda = 0.5$, 
$\theta_m = 0.25\pi$ and $\chi_m = 0.5\pi$.}
\label{fig4}
\end{figure}

A nearly similar characteristics are also seen in Fig.\ref{fig3} for a 
strongly opaque barrier with $Z_0 = 2.0$. However in this case, $Z_R = 2.0$ 
shows maximum conductance for all values of $B$. From Figs.\ref{fig3}(a) and 
\ref{fig3}(b) we have seen that though two sharp peaks still appears for 
$Z_R = 1.0$ and 
$2.0$ for low values of magnetic field $B$, but the sharpness of the peak is 
more at $E = 0.83\Delta_t$ than $E = 0.17\Delta_t$ for both $B=0$ and $B=0.3$
cases. With the rise of $B$ to 0.7 conductance get decrease but sharpness of 
the peaks retain as seen from Fig.\ref{fig3}(c). It indicate that for a 
strongly opaque barrier a moderate value $B$ is also suitable for the 
formation of ABS. For the magnetic field $B = 1.0$, a totally different 
characteristics is seen.  In this case for $Z_R = 2.0$, the spin conductance 
is maximum for zero bias condition. However, as the bias voltage
is switched on $G_S$ shows a gradual fall but shows two sharp dips exactly at 
$\Delta_\pm$. For  $0 \leq Z_R < 1$, the spin conductance spectra is quite 
similar. In all these cases, $G_S$ initially decreases monotonically from a 
maxima with the rise of $E$ and then shows a sharp minima exactly at 
$E = \Delta_\pm$. It is also seen that the sharpness of the dips are too 
strong at $E = \Delta_+ = 0.83\Delta_t$ than at $E = \Delta_- = 0.17\Delta_t$. 
\begin{figure}[hbt]
\centerline
\centerline{
\includegraphics[scale = 0.42]{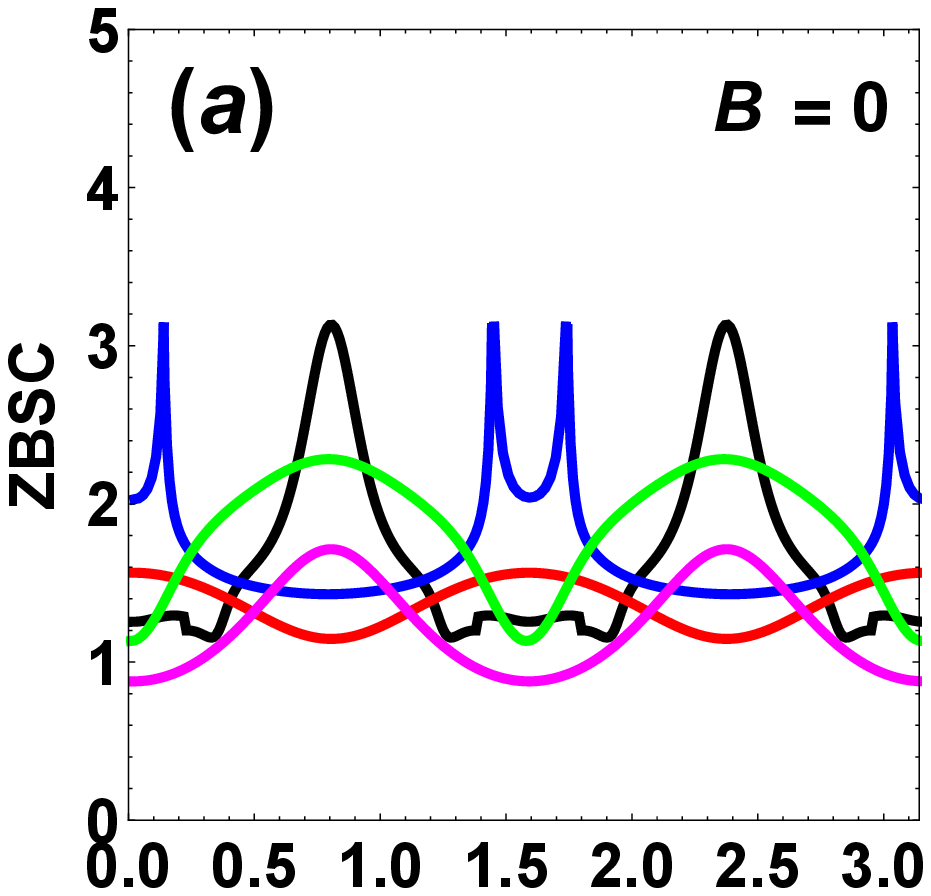}
\hspace{0.15cm}
\vspace{0.25cm}
\includegraphics[scale = 0.42]{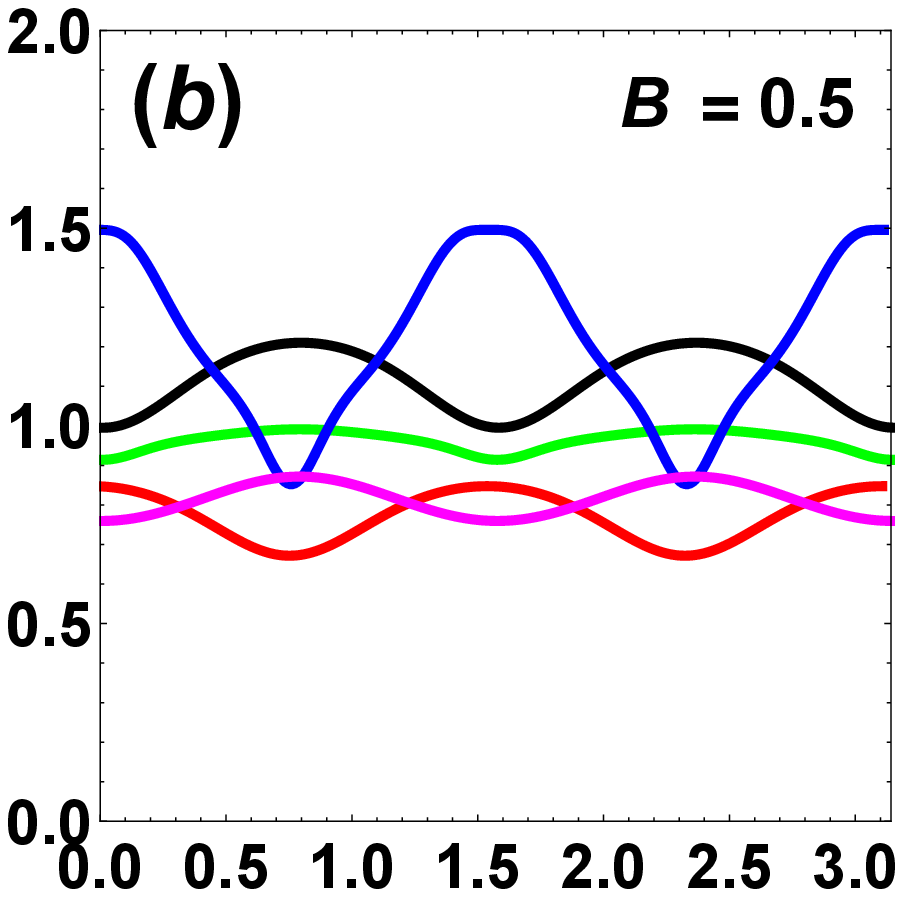}
\vspace{0.25cm}
\hspace{0.15cm}
\includegraphics[scale = 0.41]{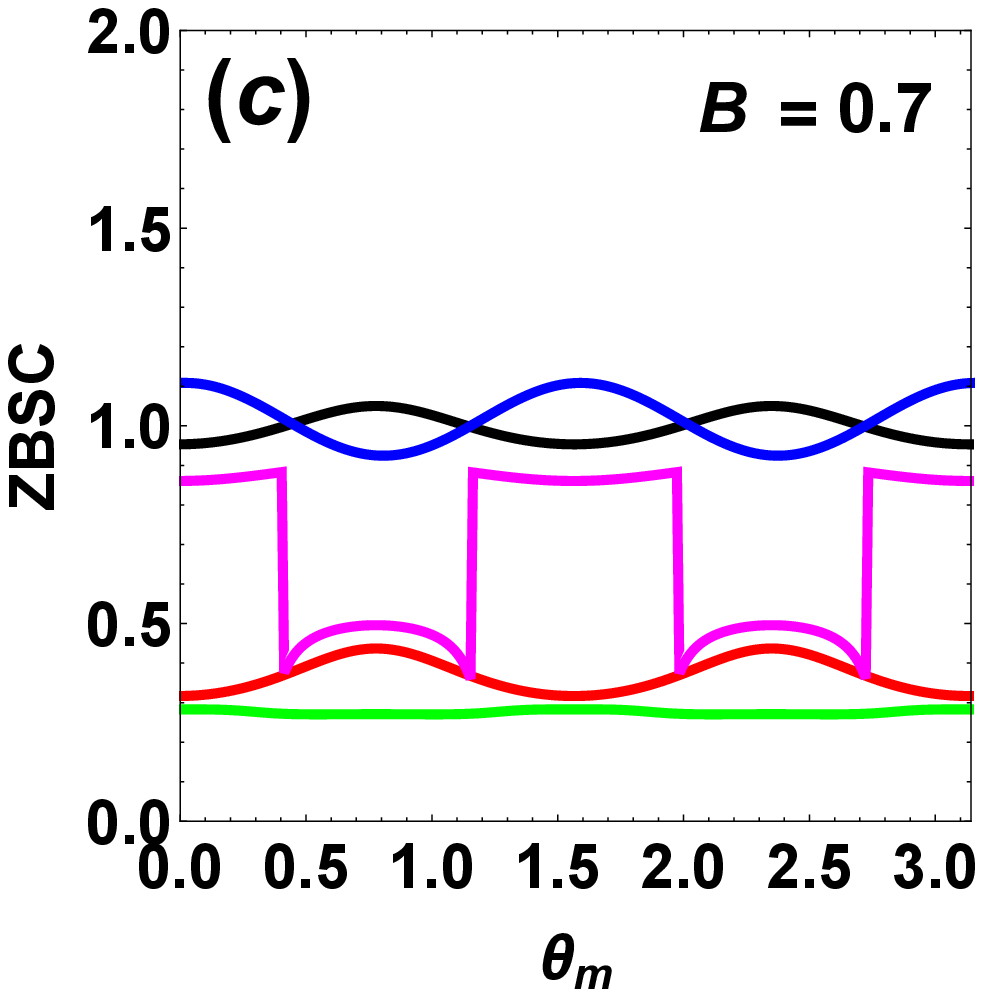}
\hspace{0.3cm}
\includegraphics[scale = 0.41]{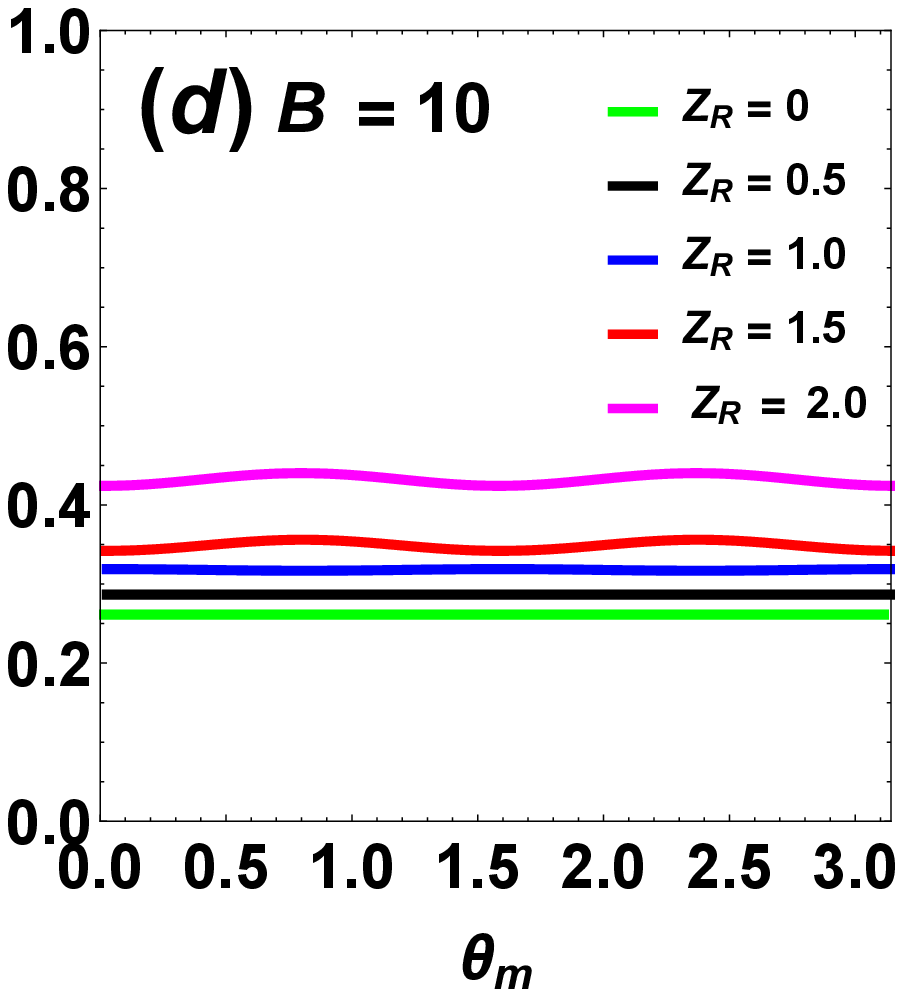}
}
\caption{Variation of ZBSC with polar angle of magnetization 
$\theta_m$ for different RSOC ($Z_R$) and in-plane magnetic field strength 
($B$). The figures are drawn for a partially transparent barrier with barrier 
width $Z_0 = 0.1$, considering azimuthal angle $\chi_m = 0.5\pi$ and FWM $\lambda = 0.5$.}
\label{fig5}
\end{figure}

\begin{figure}[hbt]
\centerline
\centerline{
\includegraphics[scale = 0.42]{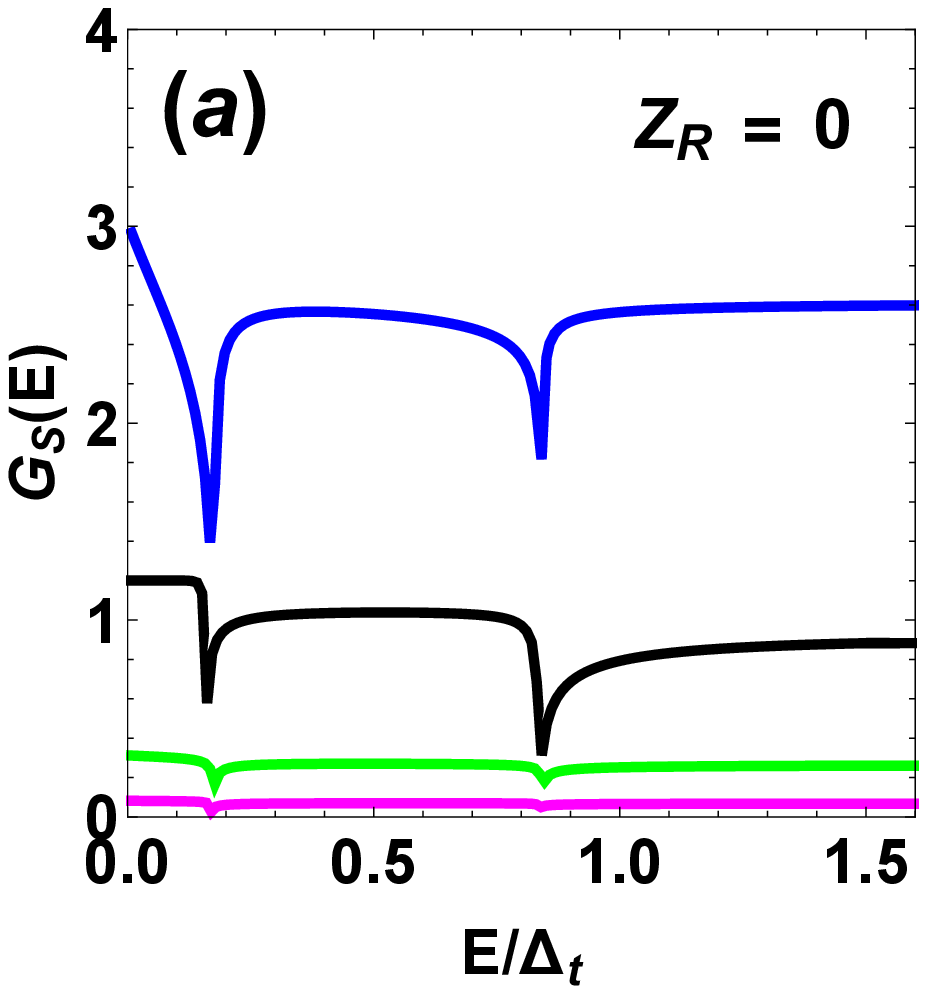}
\hspace{0.35cm}
\includegraphics[scale = 0.42]{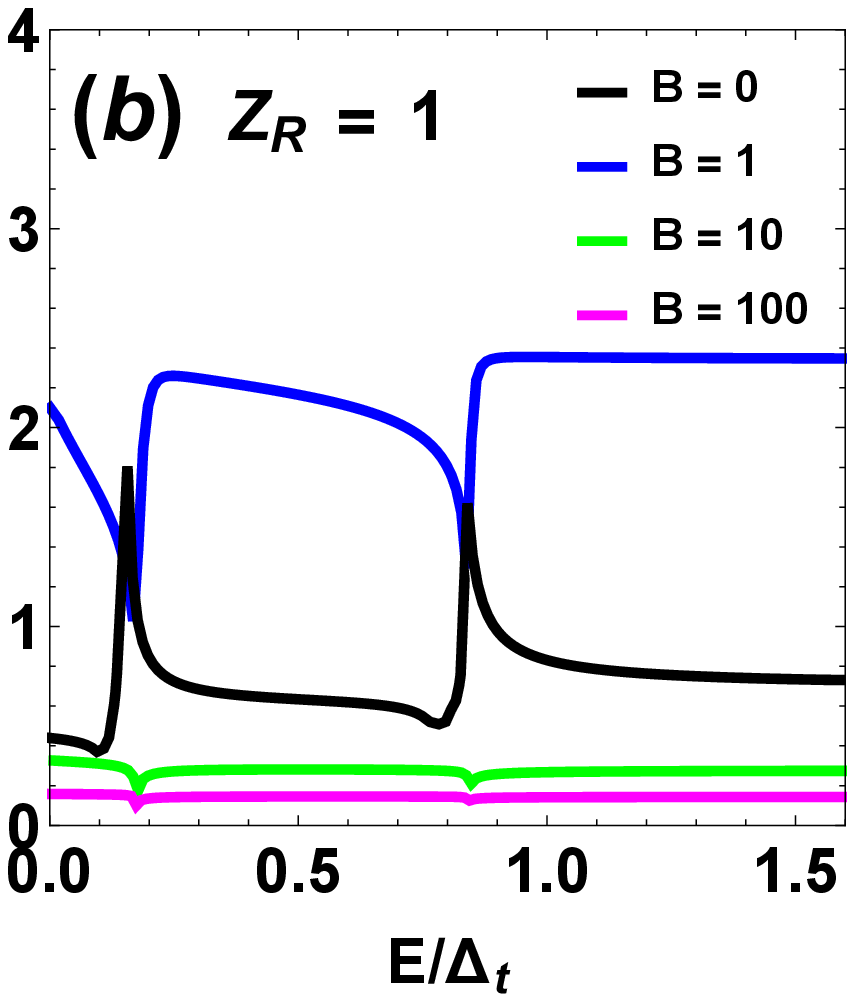}
}
\caption{Spin conductance spectra for different values of $B$ considering
$\Delta_s =\frac{\Delta_t}{3}$,  $Z_0 = 0$, $\lambda = 0.5$, $X=0.9$, 
$\theta_m = 0.5\pi$ and $\chi_m = 0.5\pi$.}
\label{fig6}
\end{figure} 

\begin{figure}[hbt]
\centerline
\centerline{
\hspace{0.05cm}
\includegraphics[scale = 0.42]{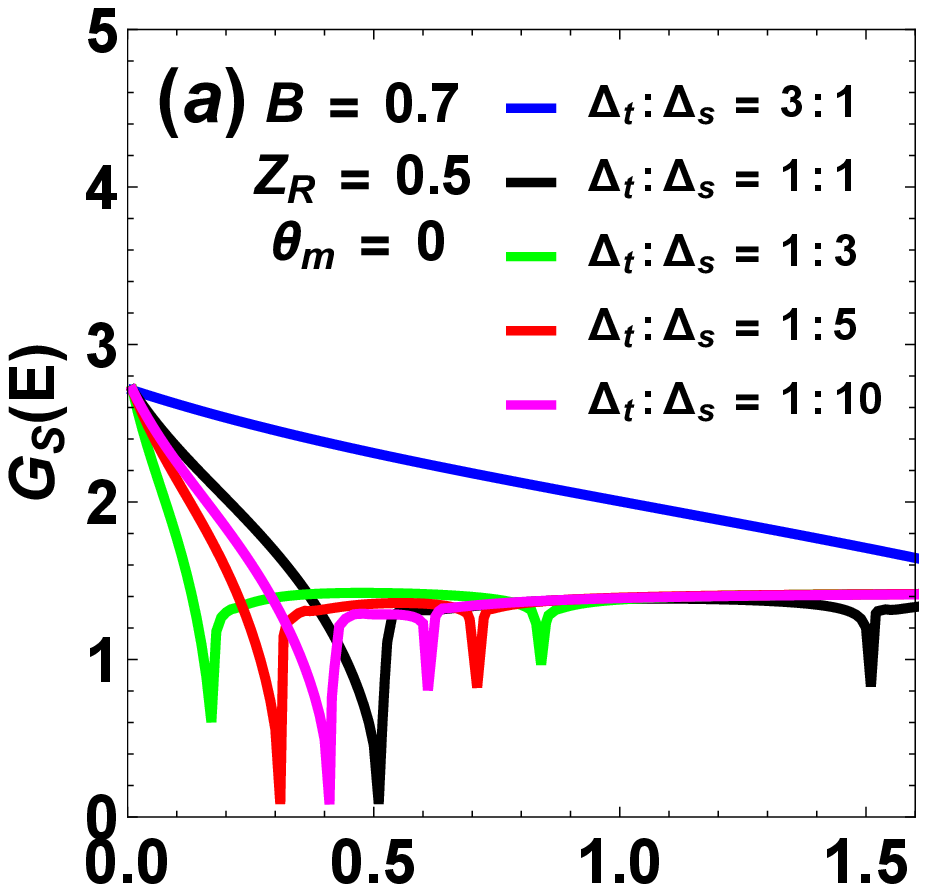}
\hspace{0.3cm}
\vspace{0.25cm}
\includegraphics[scale = 0.42]{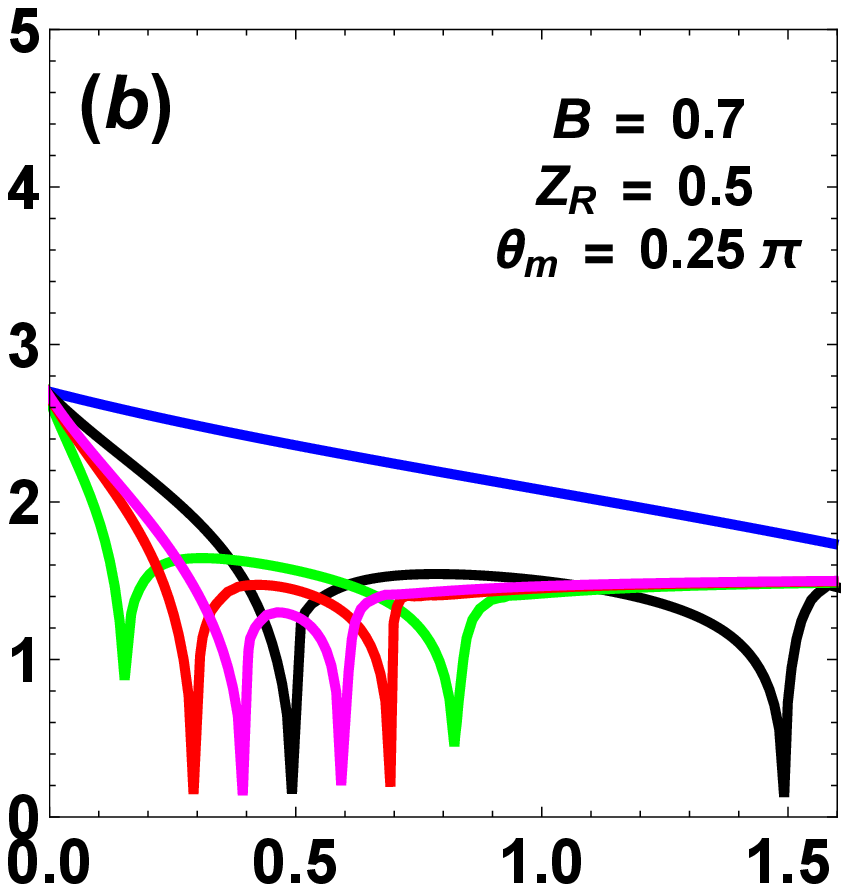}
\vspace{0.25cm}
\hspace{0.15cm}
\includegraphics[scale = 0.42]{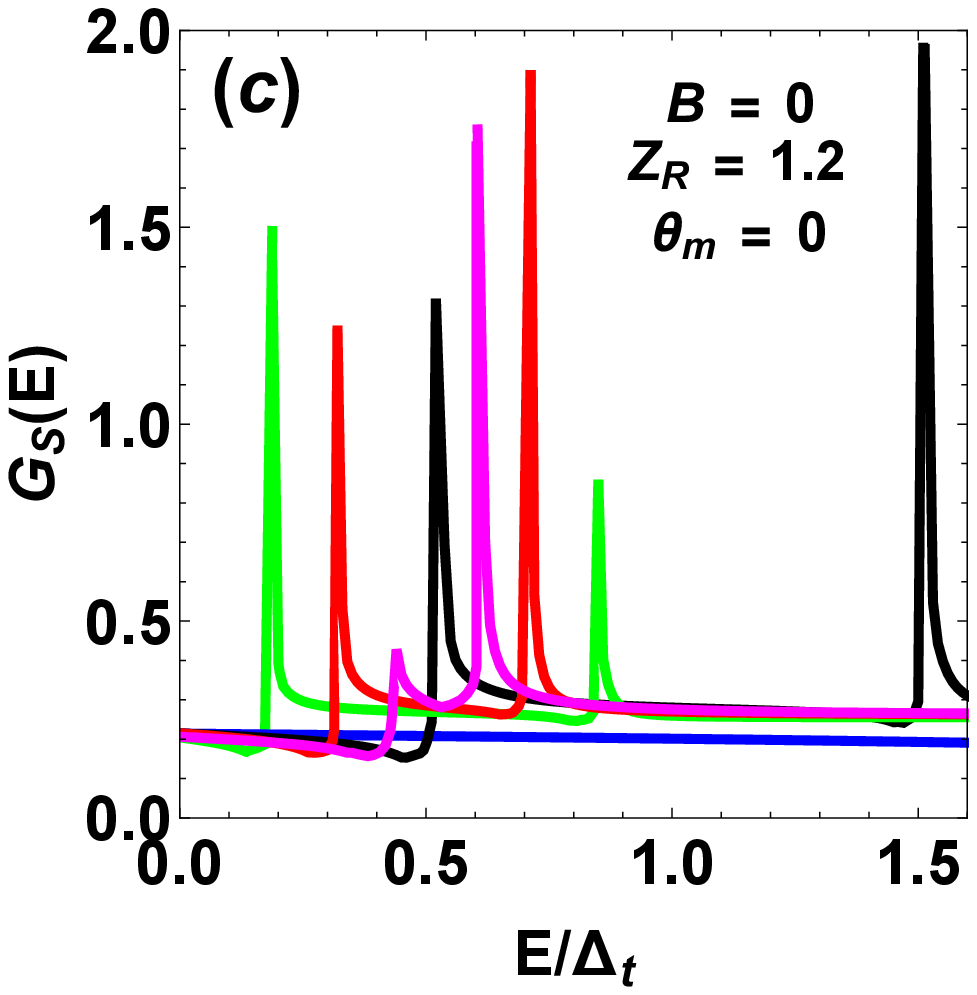}
\hspace{0.05cm}
\includegraphics[scale = 0.41]{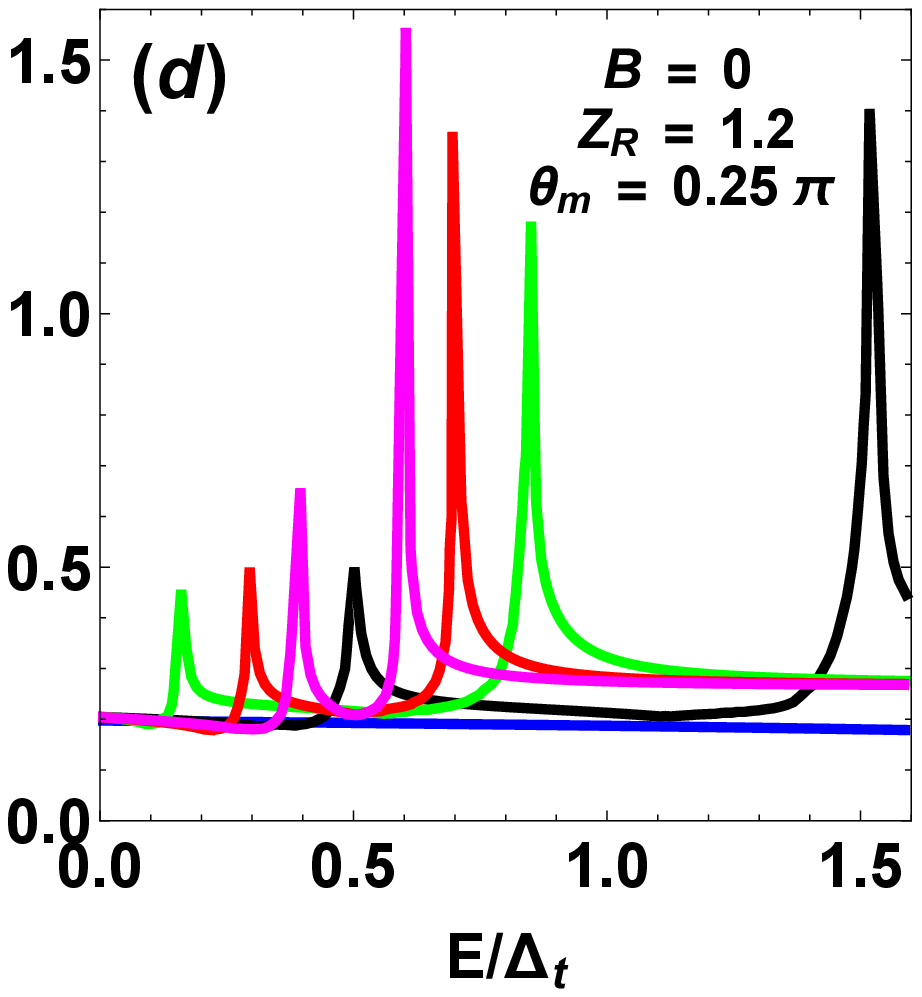}
}
\caption{Spin conductance spectra for different $\Delta_t:\Delta_s$ and $\theta_m$ 
values considering $\chi_m = 0.5\pi$,  $Z_0 = 0.1$, $\lambda = 0.5$ and 
$X=0.9$. We choose $Z_R = 0.5$, $B = 0.7$ for plots (a) and (b), 
while $Z_R = 1.2$, $B = 0$ for plots (c) and (d).}
\label{fig7}
\end{figure}

In case of a highly transparent barrier with barrier thickness $Z_0 = 0$, 
we have seen form Fig.(\ref{fig4}) that there exist a 
suppression of spin conductance from maxima 
for region $0 \leq E \leq \Delta_-$ in absence and for low $Z_R$ values. In 
case of higher $Z_R$ values i.e. $1.0$ and $2.0$, the 
conductance spectrum shows a gradual rise followed 
by a maxima at $\Delta_-$. For the region $\Delta_- \leq E\leq \Delta_+$, the 
spin conductance $G_S$ of the system shows a slow rise nearly 
at both the points $\Delta_\pm$ for $Z_R = 1.0$ and $2.0$. 
An exactly opposite characteristics is seen for $Z_R = 0$ and $0.5$. Though 
for an transparent barrier with $B = 0$, Rashba free cases shows maximum 
conductance nearly for all biasing voltages however, at $E = \Delta_\pm$, 
$Z_R = 1.0$ spectra shows maximum conductance as observed 
from Fig.\ref{fig4}(a). As soon as the magnetic field is switched on with 
$B = 1.0$, an exactly opposite characteristics is seen. In this case,
also two sharp dips are observed for $Z_R = 1.0$ and $2.0$ 
subsequently followed by two sharp peaks for biasing energy 
nearly equal to $\Delta_\pm$.  It is seen from Fig.\ref{fig4}(b) that for a 
transparent barrier with $Z_R = 0$ and $0.5$, the spin conductance spectra 
shows maximum conductance in presence of a magnetic field $B = 1.0$. Thus from 
Figs.\ref{fig2}, \ref{fig3} and \ref{fig4} it can be conclude that the 
spin conductance $G_S$ is not only dependent on barrier thickness $Z_0$, but 
the in-plane magnetic field strength $B$ and RSOC strength 
$Z_R$ play very important role in the formation of ABS in F$|$NCSC
heterostructure. It is seen that pairing and formation of ABS 
in NCSCs with moderate RSOC is suitable for a low magnetic field strength $B$. 
However, for NCSCs having moderately large RSOC and with a strongly opaque 
F$|$NCSC interface, moderate values of in-plane magnetic field is also found 
to be suitable.

It should be noted that the spin conductance is non zero even at $\theta_m = 0$
and it has a strong correlation with orientations of magnetization 
($\theta_m$, $\chi_m$), magnetization strength ($X$), the in-plane magnetic 
field ($B$) and RSOC ($Z_R$) as seen from Figs.\ref{fig2}, \ref{fig3} and 
\ref{fig4}. So, in order to understand the orientation dependence of 
$G_S$, we study the  Zero Bias Spin Conductance (ZBSC) as a function of polar 
angle of magnetization ($\theta_m$) in Fig.\ref{fig5} for different 
choices of RSOC $Z_R$ and in-plane magnetic field strengths $B$. For all our 
ZBSC spectra, we set azimuthal angle $\chi_m = 0.5\pi$, magnetization strength 
$X = 0.7$ and FWM as $\lambda = 0.5$. Moreover, we consider a partially opaque 
barrier with $Z_0 = 0.1$ for our analysis. Though many attempts had been made 
earlier to explain the ZBC in superconductors, but among them the most 
promising reason are due to phase mismatch of the transmitted ELQ particles 
and HLQ particles as reported in \cite{tanaka} and the mismatch of Fermi 
momentum in different regions (i.e. FWM) as reported in \cite{zutic1,zutic2}. 
It is seen from Figs.\ref{fig5}(a) and \ref{fig5}(b) that for all choices of 
$Z_R$,
ZBSC spectra shows an oscillatory behaviour in absence and low magnetic field 
regime, i.e. with $B = 0$ and $0.5$ respectively. A sharp Zero Bias Spin 
Conductance Dip (ZBSCD) is seen at $\theta_m = 0, 0.5\pi$ and $\pi$ for 
$Z_R = 0, 1.0$ and $2.0$ in absence of $B$, while a Zero Bias Conductance 
Peak (ZBSCP) 
is found to be observed at the same positions for $Z_R = 1.5$ as seen from 
Fig.\ref{fig5}(a). The oscillatory behaviour of the ZBSC gradually decreases 
with the rise of $B$ as seen from Figs.\ref{fig5}(b), \ref{fig5}(c) and 
\ref{fig5}(d). It is also observed from Figs. \ref{fig5}(b) and \ref{fig5}(c) 
that with the increase in $B$ from $0.5$ to $0.7$, there exist a phase 
reversal for both low $Z_R = 0$ and 
high RSOC values viz.,$1.5$ and $2.0$.
\begin{figure}[hbt]
\centerline
\centerline{
\includegraphics[scale = 0.42]{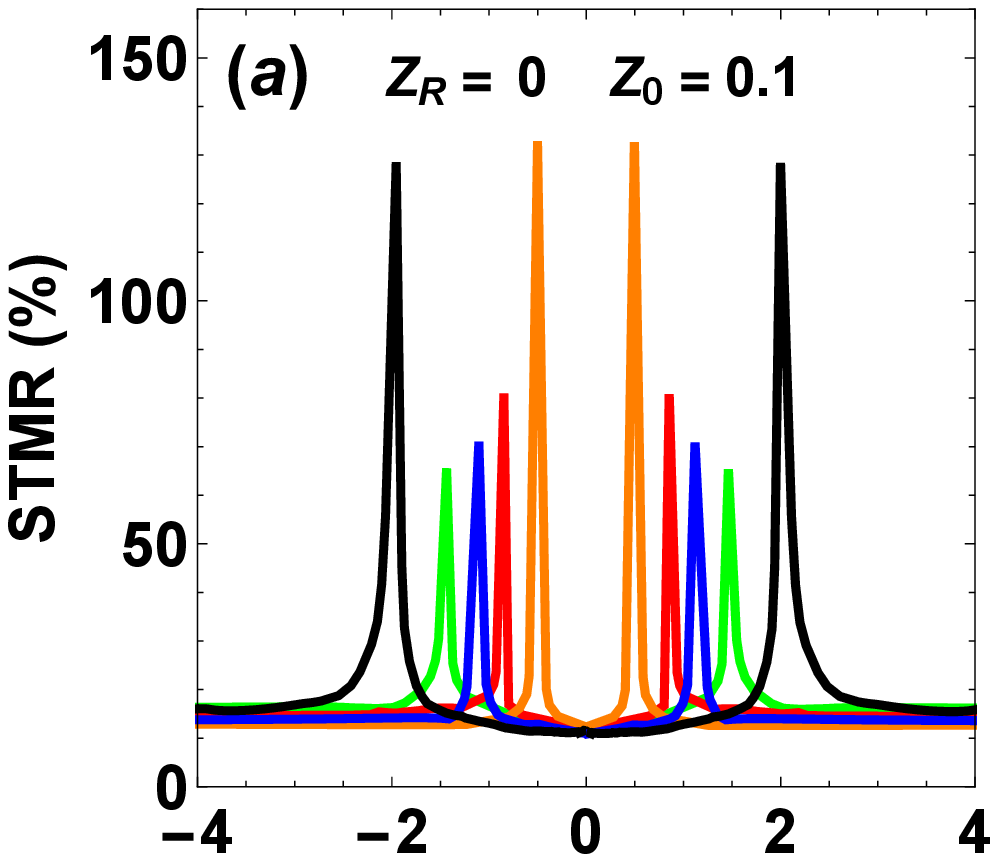}
\hspace{0.15cm}
\vspace{0.25cm}
\includegraphics[scale = 0.423]{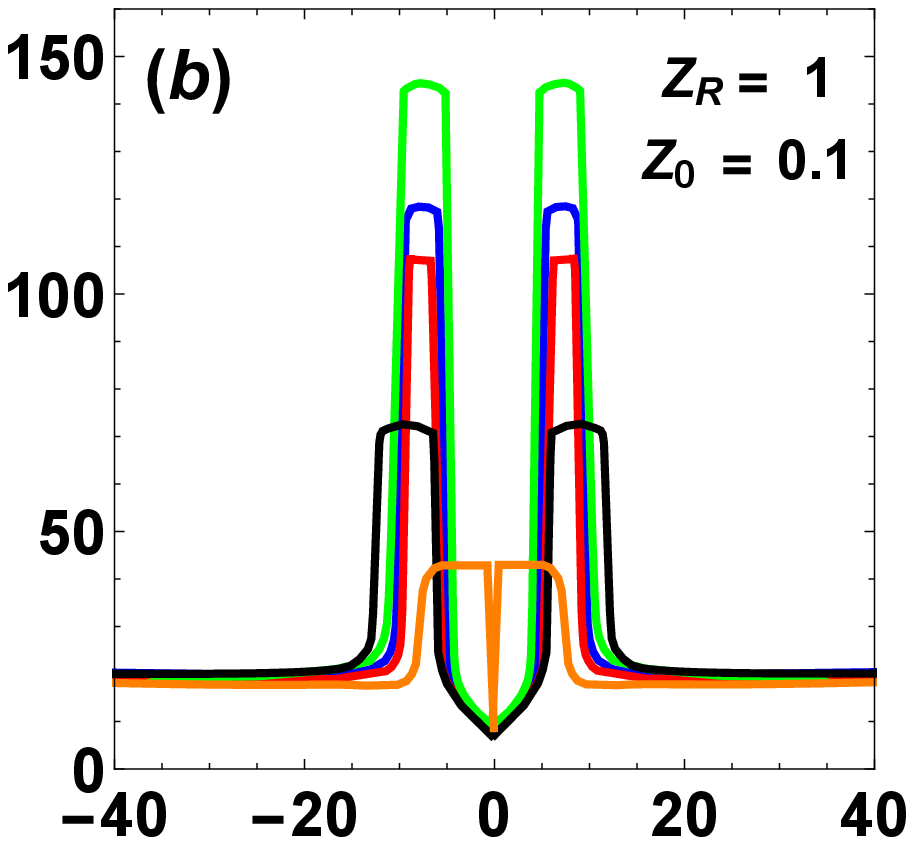}
\vspace{0.25cm}
\hspace{0.25cm}
\includegraphics[scale = 0.423]{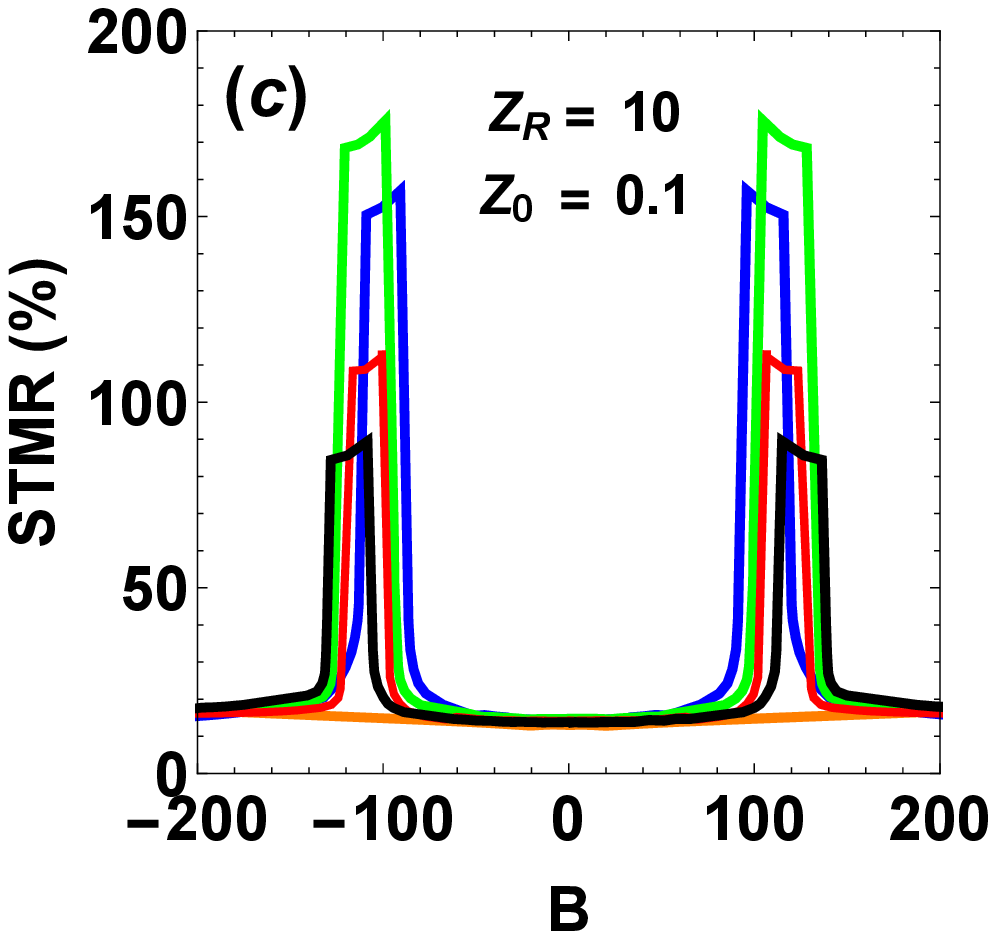}
\hspace{0.2cm}
\includegraphics[scale = 0.427]{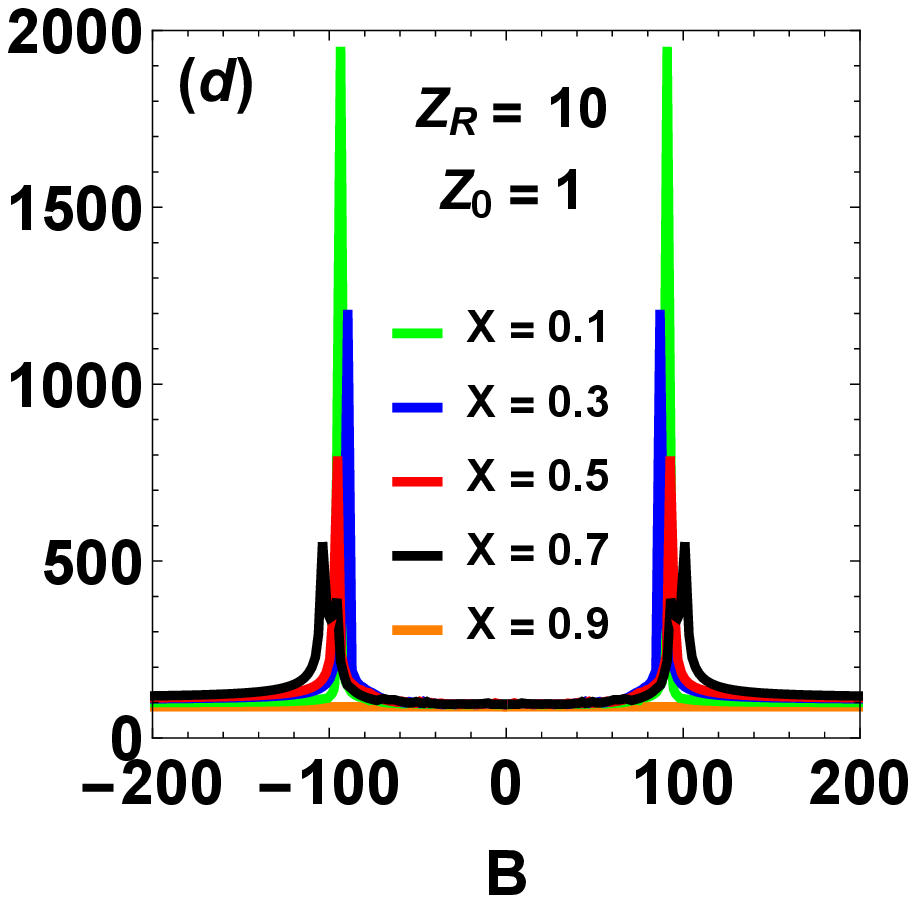}
}
\caption{Variation of STMR with magnetic field (B) for different choices of 
$X$, $Z_0$ and $Z_R$ considering
$\Delta_s =\frac{\Delta_t}{3}$ and $\lambda = 0.5$.}
\label{fig8}
\end{figure}
 The oscillatory behaviour of ZBSC 
spectra nearly die out for a very high value of $B$ to $10$ for any 
arbitrary value of $Z_R$ as seen from Fig.\ref{fig5}(d). So, it can be 
concluded 
ZBSC that has a strong dependence on the magnetization, magnetic field and 
RSOC. Notwithstanding, for an experimentally feasible design of F$|$NCSC 
heterostructure, moderate value of $B$ and NCSCs with moderate RSOC are mostly 
suitable.

\subsubsection*{Effect of in-plane magnetic field $B$}
It is already seen from the preceding section that the orientation of 
magnetization plays a significant role in the formation of ABS and pairing 
mechanism in NCSC. Moreover, the formation of spin current, STMR and 
triplet-singlet correlation in an SV are also dependent on the orientation of 
magnetization. Thus the significance of an in-plane magnetic field in the 
context of SV devices is highly inherent. So to investigate the role of 
in-plane magnetic field $B$ on spin conductance, we have studied the variation 
of $G_S$ with $E/\Delta_t$ for different choices of $B$ as shown in 
Fig.\ref{fig6}. We consider a transparent barrier with $Z_0 = 0$, 
FWM $\lambda = 0.5$, magnetization strength and orientations respectively are
$X = 0.9$, $\theta_m = 0.5\pi$ and $\chi_m = 0.5\pi$ for this study of spin 
conductance. It is to be noted here that the increasing value of $B$ 
suppresses the spin conductance of the system for any choices of RSOC. For 
Rashba free case with $Z_R = 0$, there exist two sharp dips appear at 
$\Delta_\pm$ for all choices of $B$ as seen from Fig.\ref{fig6}(a). It is 
quite obvious as the Rashba free materials doesn't offer unconventional 
superconductivity and hence the ABS are totally suppressed. However, when the 
RSOC is increased to $Z_R = 1$, two sharp peaks are observed 
exactly at $\Delta_\pm$ in absence of $B$ as seen from Fig.\ref{fig6}(b). 
It indicate the formation ABS and presence of an unconventional 
superconductivity in the SV. 
\begin{figure}[hbt]
\centerline
\centerline{
\includegraphics[scale = 0.42]{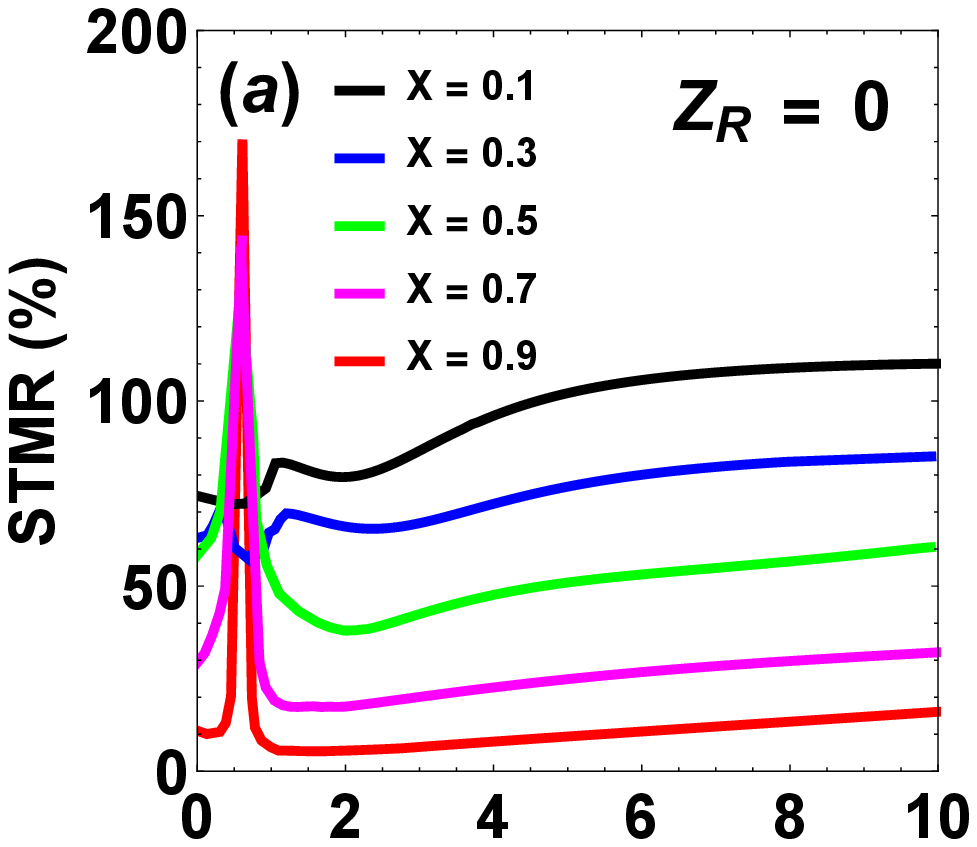}
\hspace{0.15cm}
\vspace{0.25cm}
\includegraphics[scale = 0.42]{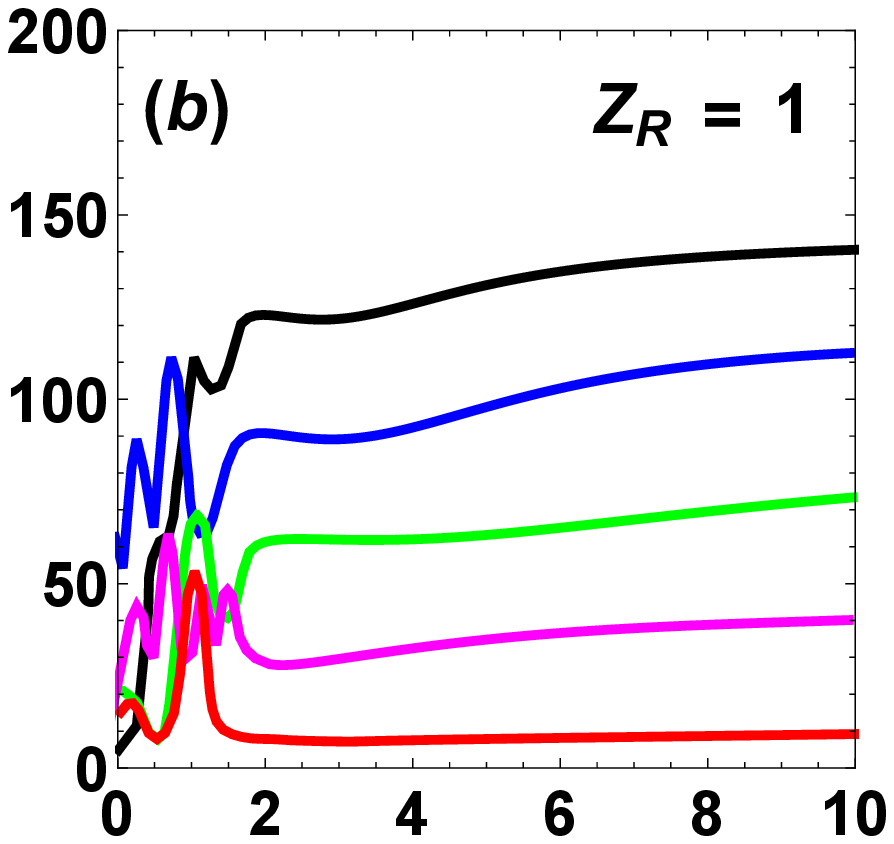}
\vspace{0.25cm}
\hspace{0.15cm}
\includegraphics[scale = 0.42]{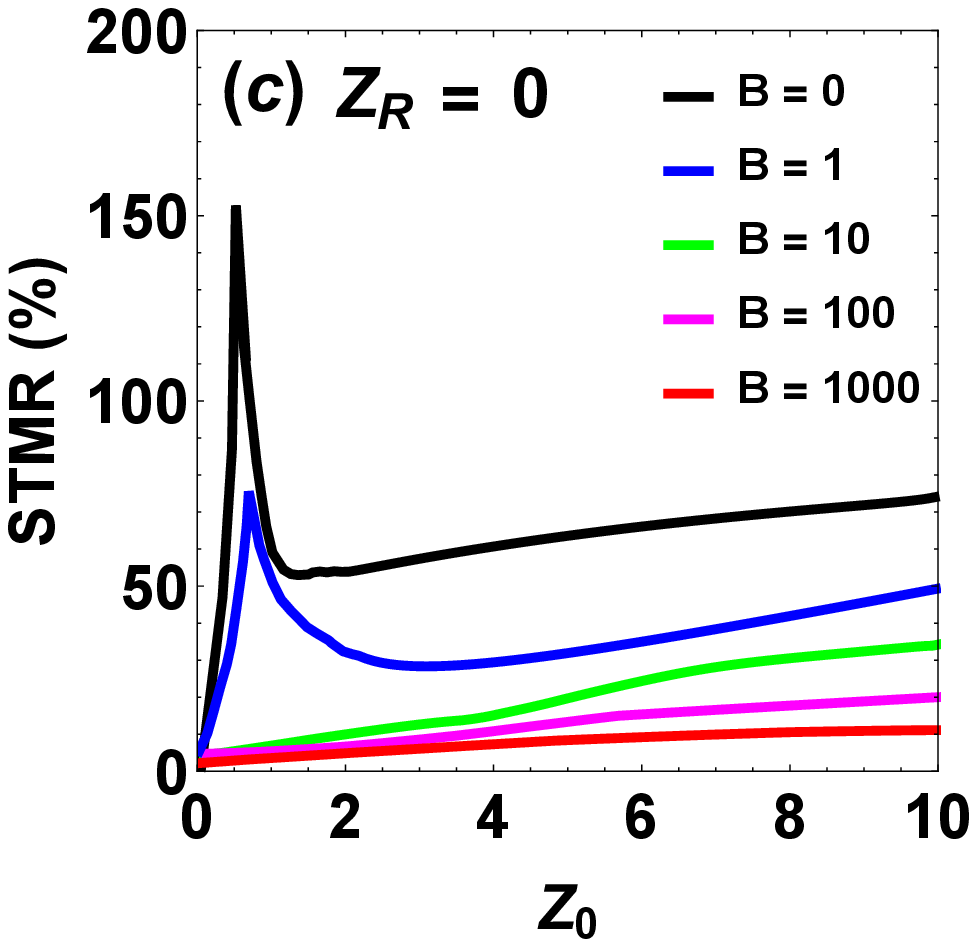}
\hspace{0.3cm}
\includegraphics[scale = 0.42]{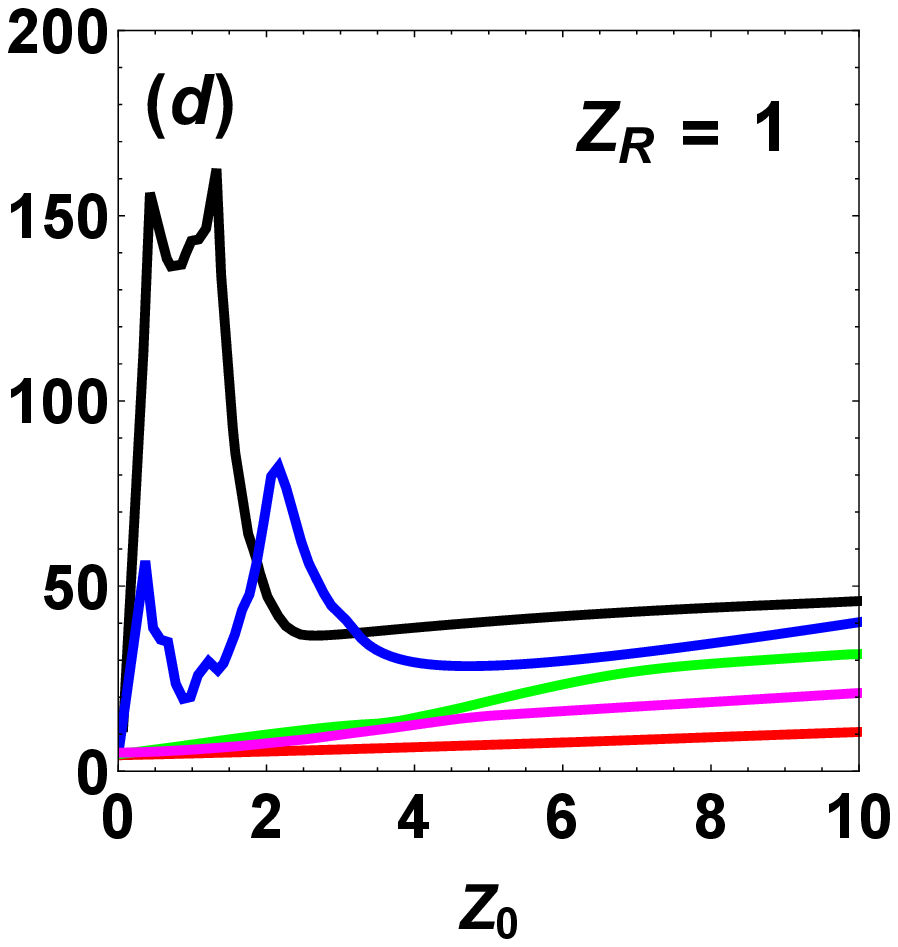}
}
\caption{Variation of STMR with barrier width $Z_0$ for different choices of 
$X$, $B$ and $Z_R$ considering $\Delta_s =\frac{\Delta_t}{3}$, 
$\theta_m = 0.25\pi$ $\chi_m = 0.5\pi$ and $\lambda = 0.5$. Figs.(a) and (b) 
are plotted for different choices of $X$ with $B = 0.1$, while Figs.(c) and 
(d) are plotted for different $B$ values with $X = 0.9$.}
\label{fig9}
\end{figure}
It is seen that if the magnetic field 
is switched on, then the pairing and the formation of ABS
will get suppressed. It is because for large magnetic field potentially 
destroy the superconducting ordering and hence the pairing mechanism. 
Thus for spin conductance and fabrication of an F$|$NCSC$|$F, NCSC materials 
with moderate RSOC is highly suitable. Moreover, though large value of 
in-plane magnetic field is not suitable as seen from Fig.\ref{fig6}, but 
for NCSC with large RSOC with opaque interface a low in-plane magnetic field 
can be preferred as already seen from Figs.\ref{fig2} and \ref{fig3}.

\begin{figure*}[hbt]
\centerline
\centerline{
\includegraphics[scale = 0.48]{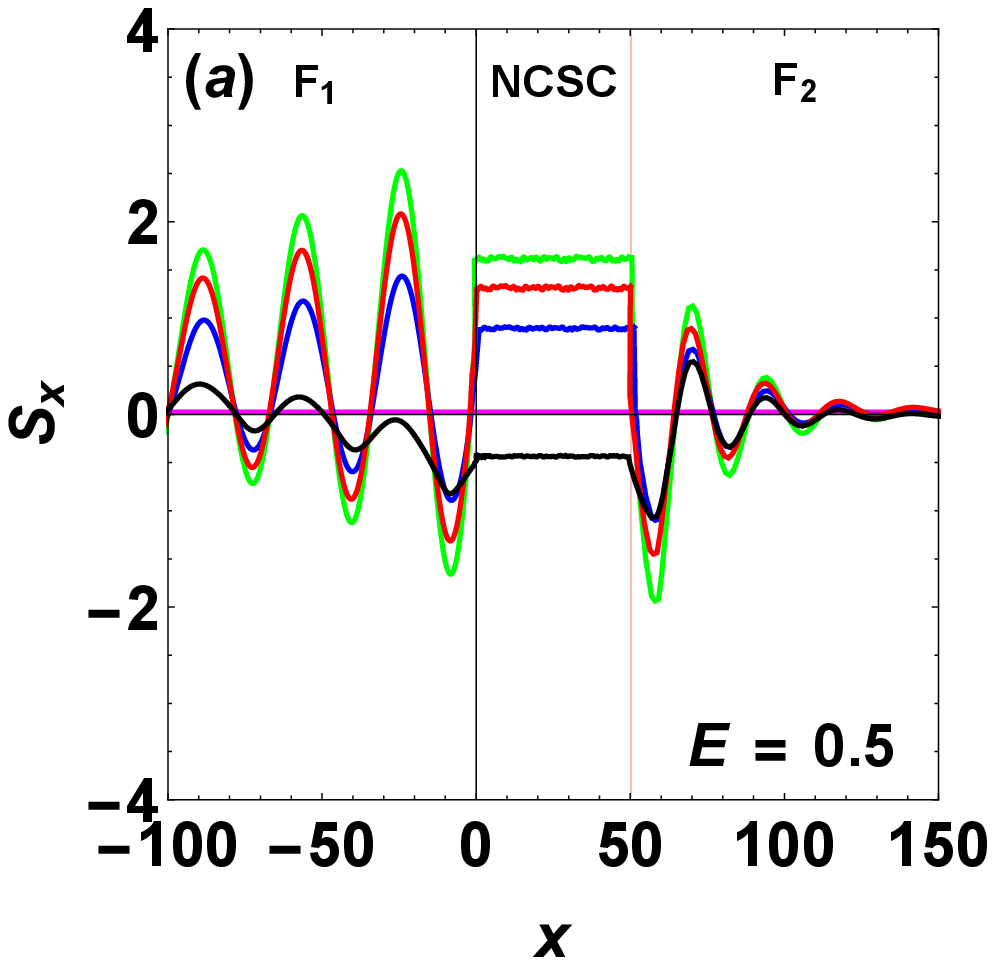}
\vspace{0.35cm}
\hspace{0.4cm}
\includegraphics[scale = 0.48]{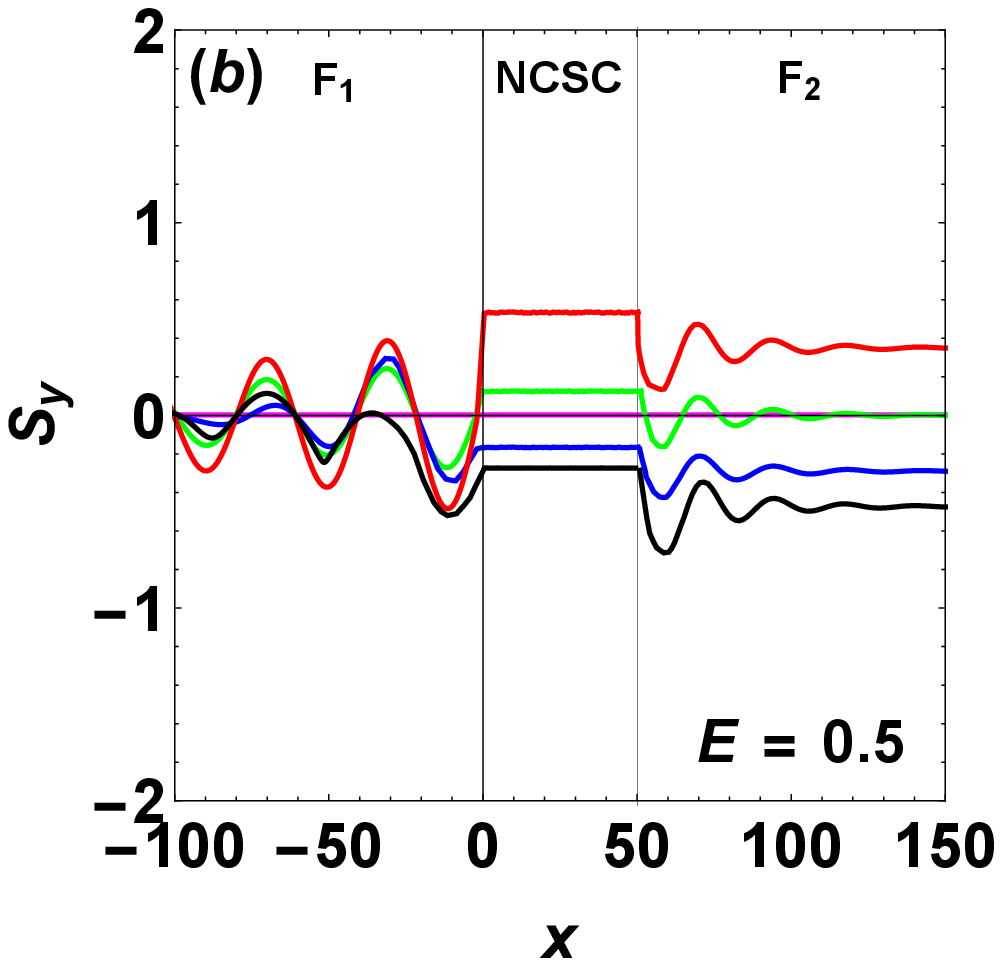}
\vspace{0.35cm}
\hspace{0.4cm}
\includegraphics[scale = 0.48]{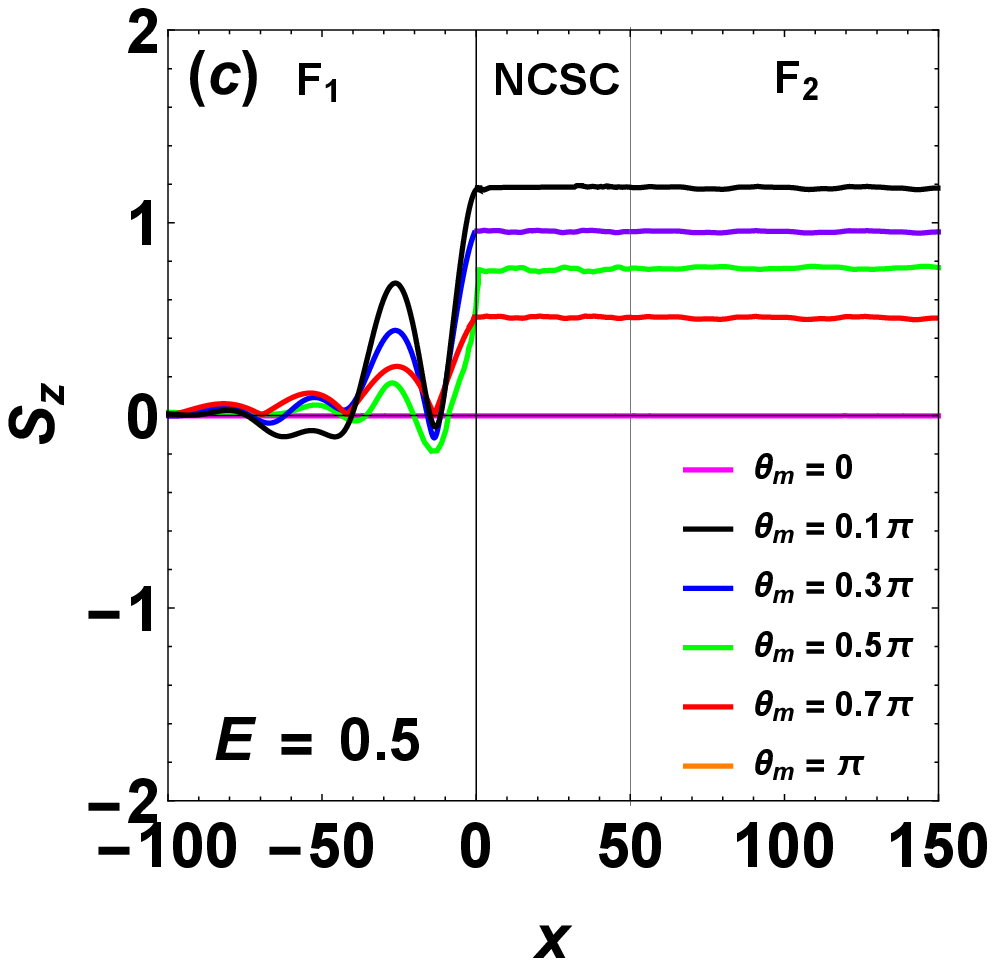}
\vspace{0.35cm}
\includegraphics[scale = 0.48]{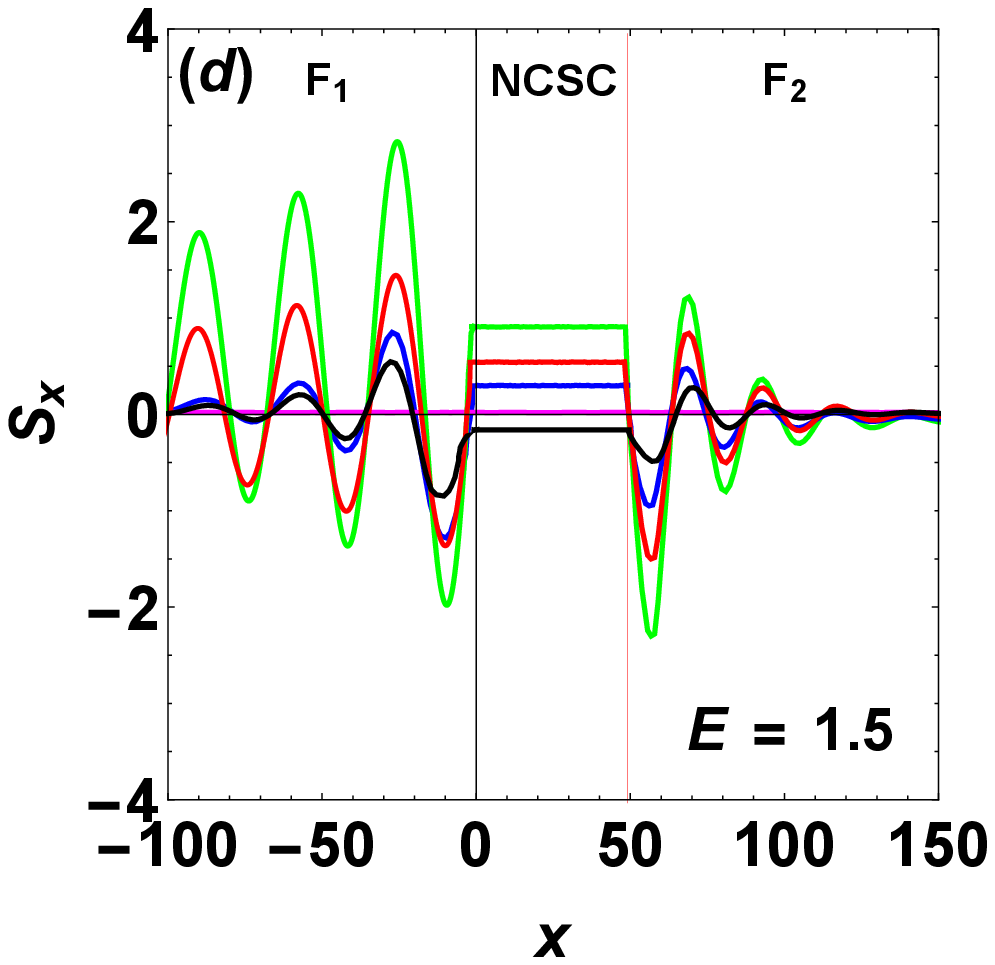}
\vspace{0.35cm}
\hspace{0.4cm}
\includegraphics[scale = 0.49]{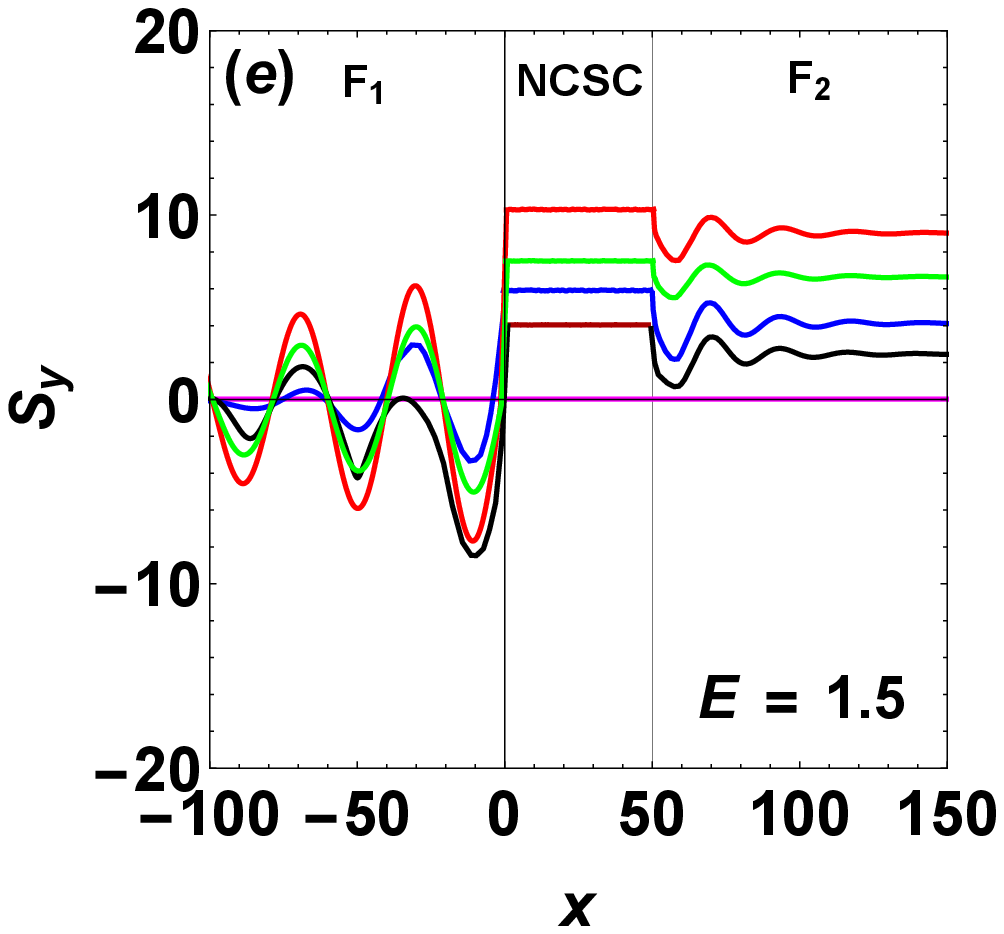}
\vspace{0.35cm}
\hspace{0.4cm}
\includegraphics[scale = 0.48]{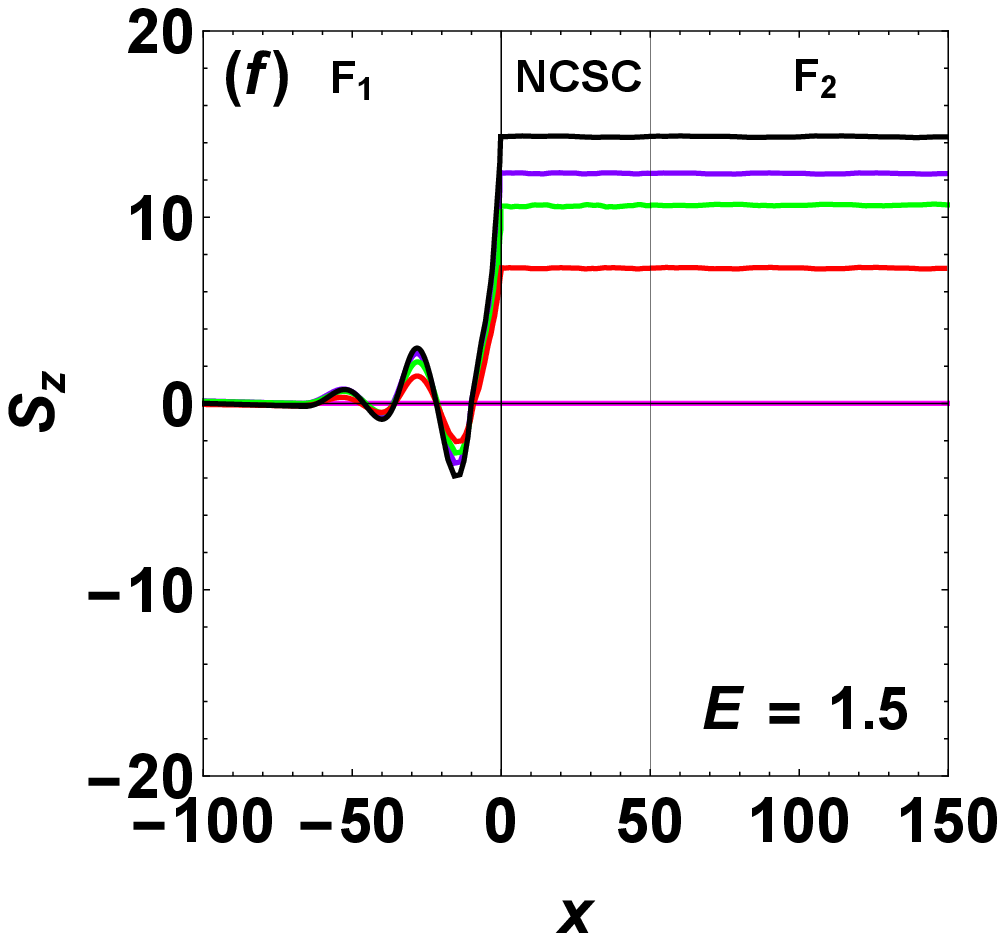}
\vspace{0.35cm}
}
\caption{Spatial variation of spin current for two different bias voltages 
viz., $E = 0.5$ and $1.5$. The components of spin current 
($S_x, S_y, S_z$) are plotted in y-axes are the multiple of $10^{-4}$, while 
the spatial coordinate ($x$) is taken in nanometre scale. The figures are 
plotted for $\Delta_s =\frac{\Delta_t}{3}$, $\lambda = 0.5$, $Z_0 = 0.1$, 
$Z_R = 1$, $X = 0.9$ and $\chi_m = 0.5\pi$.}
\label{fig10}
\end{figure*}

\subsubsection*{Effect of singlet-triplet mixing ratio $\Delta_s:\Delta_t$}
It is of our interest to study the role of pairing amplitude on spin 
conductance. So, in Fig.\ref{fig7} we study the variation of spin 
conductance $G_S$ with $E/\Delta_t$ for different choices of singlet-triplet 
mixing ratios. For our analysis, we consider a partially transparent 
barrier with barrier thickness $Z_0 = 0.1$, FWM $\lambda = 0.5$, 
magnetization strength $X = 0.9$ and the azimuthal angle $\chi_m = 0.5\pi$. 
Furthermore, we consider $B = 0.7$ and $Z_R = 0.5$ for our 
Figs.\ref{fig7}(a) and \ref{fig7}(b). It is seen that the spin conductance has 
been concealed 
at $\Delta_\pm$ for $\Delta_t\leq\Delta_s$ in both the cases as already
seen in Figs.\ref{fig2}, \ref{fig3}, \ref{fig4} and \ref{fig6} 
too. It is to be noted that if $\Delta_t>\Delta_s$, the 
spin conductance falls linearly which indicate that the absence 
of ABS in such materials. It is also seen that though triplet correlation 
is favoured in many NCSCs however, the percentage of singlet correlation 
is always greater than the triplet correlation in NCSC. It is also observed 
that the sharpness of the dips increases with the change in $\theta_m$ from 
$0$ to $0.25\pi$. Beside that we also noted that for $\theta_m  = 0$ 
$E =\Delta_-$ provides maximum suppression, while for $\theta_m  = 0.25\pi$ 
$E =\Delta_+$ provides the same. Though ABS is seen in 
Figs.\ref{fig7}(a) and \ref{fig7}(b) however as the magnetic field is switched 
off and RSOC is increased to $Z_R = 1.2$, superconducting behaviour is 
achieved again. 
The singlet-triplet correlation in such cases is very significant 
and are strongly favourable for the formation of ABS at $\Delta_\pm$ as seen  
in Figs.\ref{fig7}(c) and \ref{fig7}(d). Moreover, it is also observed that 
with the 
increase in the singlet components the ABS will approach to each other and 
also found to be observed at lesser bias voltages. So from the Fig.\ref{fig7} 
it can be concluded that ABS can be tuned by singlet-triplet pairing 
ratio, in-plane magnetic field and also by RSOC. 
For an experimentally suitable scenario, NCSC with greater singlet 
components than triplet are mostly suitable. Moreover, arbitrary orientation 
of magnetization is mostly suitable for fabrication purpose. 

\subsection{Spin Tunnelling Magneto-Resistance (STMR)}
Over the last two decades, MTJs have gained attention owing to its robust 
physics along with the potential applications in spintronic devices. 
The discovery of giant Tunnelling Magneto-Resistance
(TMR) \cite{binasch,baibich,parkin,chappert} in MgO based MTJ 
is one of the key reason behind its increased application in 
Magnetic Random Access Memory (MRAM) and magnetic sensors. The 
study of TMR reveals many novel properties about F-N and F-S based MTJs over 
the years. These studies mainly involves magnetic response of MTJ, spin 
alignments, dependence on temperature and barrier width. Introduction of an 
NCSC as a spacer can enhance the TMR value and hence found to be useful to 
fabricate ultra fast cryogenic MRAM. The basic reason to consider NCSC as a 
spacer is because of the existence of the lack of inversion symmetry and hence 
it shows unconventional superconductivity and possess a strong RSOC. 
Moreover, they can also support flow of polarized current.
In view of this, we have studied the variation of STMR ($\%$) 
as a function of in-plane magnetic field ($B$) in Fig.\ref{fig8}
and barrier width ($Z_0$) in Fig.\ref{fig9} for different magnetization 
strengths $X$, RSOC $Z_R$. We again set, $\Delta_s =\frac{\Delta_t}{3}$, 
$\lambda = 0.5$ and a consider nearly transparent barrier with $Z_0 = 0.1$ in 
Figs.\ref{fig8}(a), \ref{fig8}(b) and \ref{fig8}(c), while in Fig.\ref{fig8}(d) we choose a 
partially opaque barrier with $Z_0 = 1$. In all the cases two sharp symmetric peaks are observed which arises due to the opposite
alignment of the spins. For Rashba free case $Z_R = 0$ with a highly transparent barrier $Z_0 = 0.1$, these peaks are observe for 
a significantly low magnetic field. However, it is seen that STMR value decayed too rapidly for large value of $B$ as seen from Fig.\ref{fig8}(a). It is
to be noted here that the STMR increases with the increase in $X$ and becomes
maximum for $M = 0.9E_{FF}$ in this case.
With the rise of $Z_R$ to 1, the peaks dissappear and two flat
region are observed for a 
moderately strong field as seen from Fig.\ref{fig8}(b). It is due to the 
enhancement of opposite spin correlation with RSOC. A similar scenario
is also observed from Fig.\ref{fig8}(c) for $Z_R = 10$.
However, in this case the flat
regions appear at a very strong magnetic field $B$. For, a partially
opaque barrier ($Z_0 = 1$) with $Z_R = 10$, the flatness disappears and
two sharp peaks reappear for large values of $B$ as seen
from Fig.\ref{fig8}(d). In this case a very 
giant STMR value ($\sim 2000\%$) is observed for a magnetization
strength of $M = 0.1E_F$ at a magnetic field
$B$ $\sim 100$. It is also seen from Fig.\ref{fig8}(d) that the STMR $(\%)$
drastically 
reduced as $M\rightarrow E_{FF}$ and it becomes minimum
for $M = 0.9E_{FF}$.
It is to be noted that the STMR value reduces to zero for $M = E_{FF}$ as seen
earlier from Eqs.(\ref{eq18}) and (\ref{eq19}).
An exactly opposite scenario is seen from Fig.\ref{fig8}(a) for a highly 
transparent barrier ($Z_0 = 0.1$) with $Z_R = 0$. In this 
case for $M = 0.9E_{FF}$ STMR $(\%)$ is found to be maximum. Thus it can be
inferred that for the fabrication of MTJs with NCSC having large RSOC, moderate
magnetization strength can be found to be useful. However, a very strong
magnetic field is required for this purpose. It can also be concluded that with the increase in barrier width the STMR($\%$) increases.
Hence, NCSC's with moderate or strong RSOC and F having low exchange energy
with partially opaque barrier between them is highly preferable for
development of an MTJ with NCSC as a spacer.  

We have seen from above that the role of barrier width $Z_0$
is too inherent to determine the STMR value of an MTJ.
Thus to understand the role of $Z_0$ completely
we have also studied the variation of STMR ($\%$) with 
$Z_0$ in Fig.\ref{fig9} for different choices of $Z_R$, $X$ and $B$.
More specifically in Figs.\ref{fig9}(a) and \ref{fig9}(b), we have studied the
variation of STMR ($\%$) with $Z_0$ for different 
choices of $X$ having a fixed magnetic field $B = 0.1$, while in
Figs.\ref{fig9}(c) and \ref{fig9}(d) we have considered different $B$ values
with a fixed magnetization strength $X = 0.9$. It is seen that for Rashba free 
case with the increase in $Z_0$, the STMR $(\%)$ sharply increases initially 
for all $0.5 \le X < 1.0$ and reaches a
maximum at $Z_0\sim 0.5$. With the further rise in $Z_0$, the STMR
value shows a sharp decrease followed by a gradual rise and
saturates at large values of $Z_0$ in all cases of $X$ as seen from
Fig.\ref{fig9}(a).
It is to be noted here that for $X = 0.1$ and $0.3$,
STMR value is found to be too low and no significant peaks are observed
which is in accordance with the result found in Fig.\ref{fig8}(a). 
A totally different characteristics has been
observed from Fig.\ref{fig9}(b) in low $Z_0$ regions with
RSOC is increased to $Z_R = 1$.
In this case with the increase in $Z_0$, the STMR value 
initially fluctuate and gradually saturates for $Z_0 > 2$ for all choices of 
$X$. It is to be noted here that in both the cases, $M = 0.1E_{FF}$ shows
maximum STMR value for large values of $Z_0$ as seen from Figs.\ref{fig9}(a) 
and \ref{fig9}(b). 
A similar peak is also 
observed nearly at the same position in Rashba
free case $Z_R = 0$ with $B = 0$ and $1$, as seen from 
Fig.\ref{fig9}(c). It is to be noted that for a moderate
and strong field $B$, the peak gets disappeared and STMR value 
shows a linear rise and saturates with the increase in $Z_0$.
The STMR spectra shows a quite similar behaviour for $Z_R = 1$ with 
$B = 0$ and $1$ also as seen
from Fig.\ref{fig9}(d). However, in this case STMR value gets reduced
for large value of $B$. It is to be noted that there exist two sharp
peaks for $Z_0 < 2$ with $B = 0$, which is in accordance with our 
Fig.\ref{fig8}(b). A similar pattern is also seen for $B = 1$. However for this 
case the STMR value becomes maximum for $Z_0 \sim 2$.
Moreover, from Figs.\ref{fig9}(a) a large STMR 
value of $\sim 170\%$ is seen 
for $X = 0.9$ with $B = 0.1$. It is also to be noted that this
STMR value decreases to $\sim 50\% $ for $Z_R = 1$ as
seen from  Figs.\ref{fig9}(b).
As soon as the magnetic field is switched off, the STMR value
is found to be $\sim 150\%$ in absence of RSOC as
seen from Fig.\ref{fig9}(c). However, for $Z_R = 1$ with $B = 0$,
the STMR value is slightly reduced.
It is to be noted that though the STMR ($\%$) value is 
($\sim 170\%$) in Rashba free case, but for $Z_R = 1$ it 
decreases to ($\sim 50\%$) for $X = 0.9$ seen from Fig.\ref{fig9}(b). 
As it is seen that there exist two sharp peaks for $B = 0$ and $1$, 
however for large value of $B$ the peaks disappeared again and
no significant change is seen from Rashba free case. 

\subsection{Spin Current}
To understand the role of exchange coupling and its interplay 
with the external bias voltages and in-plane magnetic filed applied to  
the proposed system, we have examined the behaviour of the spin 
current that exist in the SV. More specifically, 
we have investigated the components of spin current ($S_x, S_y, S_z$) 
as a function of the spatial coordinate $x$ at low bias ($E = 0.5$) 
in top panel and high bias ($E = 1.5$) in bottom panel of Fig.\ref{fig10}. 
The positions of the interfaces are indicated by the vertical line with 
origin is chosen to be the F$|$NCSC interface. In our setup, the left 
F$_1$-layer represent a soft ferromagnet with exchange field 
$\boldsymbol{h}_1 = h_0(\sin\theta_m \cos\chi_m, \sin\theta_m\sin\chi_m, 
\cos\theta_m)$, while the right F$_2$-layer represent a hard ferromagnet with 
exchange field $\boldsymbol{h}_2 = h_0(0, 0, 1)$ as already mentioned above. 
For our analysis we set, 
$\Delta_s =\frac{\Delta_t}{3}$, $\lambda = 0.5$, $X = 0.9$, $\chi_m = 0.5\pi$, 
$Z_0 = 0.1$ and $Z_R = 1$. Furthermore, we have also considered six different 
choices of the polar angle of magnetization $\theta_m$ ranging from
$0$ to $\pi$ in each panel of Fig.\ref{fig10}. It is observed that for 
parallel ($\theta_m = 0$) and anti parallel($\theta_m = \pi$) orientations of 
the magnetizations the spin current vanishes, which is quite obvious 
because of the vanishing STT. It is also seen that if the polar angle 
$\theta_m$ of exchange field is slightly rotated to an angle $0.1\pi$, a 
negative spin current flows in the F$_1$ and NCSC region.
With the further rotation of $\theta_m$ the spin current reverses 
its polarization and becomes maximum for the orthogonal configuration as seen 
from the Fig.\ref{fig10}(a). A similar behaviour is also observed 
earlier in Ref\cite{moen2,halterman5,halterman6,linder5}. If $\theta_m$ 
is rotated further to $0.7\pi$ the spin current decreases. Since the spin 
current is conserved in the NCSC region all the components of $\boldsymbol{S}$ 
remains constant. In the hard ferromagnet since the exchange field is along 
the z-direction, it is observed that the transverse components $S_x$ and 
$S_y$ decayed too rapidly, while the longitudinal component $S_z$ remains 
constant as seen from  Fig.\ref{fig10}. It is to be noted here that with the 
increase in bias voltage $E$ to $1.5$ the $S_y$ and $S_z$ spin current 
increases, however $S_x$ nearly remains same  as seen from 
Figs.\ref{fig10}(a) and \ref{fig10}(d). It is because $S_x$ is primarily 
driven by the spin torque which exists. It is to be noted that the $S_y$ 
component of spin current is opposite in phase with $S_x$ and $S_z$ in both 
biasing situation. For low biasing 
it becomes negative for angle $\theta_m \le 0.3\pi$ and decayed for 
orthogonal orientation. However, for a high biasing $S_y$ component becomes 
positive for all orientations as seen from Figs.\ref{fig10}(b) and 
\ref{fig10}(e). It is also observed that 
the $S_z$ component of the spin current increases very rapidly in the 
F$_1$ region for high biasing as seen from Figs.\ref{fig10}(c) and \ref{fig10}(f).
 
\section{Conclusions}
In summary, in this paper we have investigated the spin transport in the 
F$|$NCSC$|$F SV. More specifically, the TSC, STMR and the spin current have 
been studied for an experimentally suitable parameter set of the proposed SV.
To study the normalized TSC at the F$|$NCSC interface we use an extended 
BTK approach and the scattering matrix formalism. We consider a Rashba type 
spin orbit coupling (RSOC), different strength, alignments of the exchange 
field and also for different singlet triplet mixing ratios of the gap 
amplitudes to study spin conductance. Furthermore, we consider an in-plane 
external magnetic field to develop the BdG Hamiltonain. We also consider a 
experimentally suitable value of the FWM for our investigation of spin 
conductance, STMR and spin current. We have also studied the STMR and spin 
current for different orientation of exchange field. Our results reveals many 
useful information of the F$|$NCSC$|$F SV system. It is seen that the spin 
conductance has a strong dependence on RSOC, barrier width, singlet-triplet 
correlation, in-plane magnetic field and magnetization. For a SV with nearly 
transparent barrier, NCSCs with moderate RSOC show large conductance in 
absence of magnetic field. It is seen that with the rise of magnetic field the 
spin conductance and the formation of of ABS gets suppressed. However, for 
partially and strongly opaque barriers, a very low and moderate value of 
magnetic field is suitable for formation of ABS. Beside that it can also be 
concluded that ABS can be tuned by singlet-triplet pairing ratio, in-plane 
magnetic field and also by RSOC. For fabrication of superconducting spintronic 
device, NCSC materials with more singlet components than triplet are mostly 
found to be suitable. There exist a ZBSCP and a ZBSCD which strongly indicate 
that spin conductance is orientation dependent. 
In addition, we have seen a significantly large STMR ($\%$) for the 
proposed setup. It is found that for fabrication of superconducting MTJs with 
opaque barrier, NCSC with large RSOC and high magnetization strength is highly 
suitable. But for a nearly transparent and partially opaque barrier, NCSCs 
with moderate RSOC, low magnetic field and low values of magnetization 
strength is strongly preferred. Moreover, from the study of spin current we 
have seen that it is strongly orientation dependent. With the increase in 
bias voltage spin current increases in traverse direction but the component 
along the direction of flow is almost independent. 

We sincerely hope that the results of our work on F$|$NCSC$|$S SV will shed 
some light in the field of superconducting spintronics which can be utilized 
to fabricate practical devices in near future.


\begin{thebibliography}{99}
\bibitem{johnson}
M. Johnson, Phy. Rev. Lett. {\bf 70}, 2142 (1993).
\bibitem{jedema}
F. J. Jedema, A. T. Filip and B.J.van Wees, Nature (London), {\bf 410}, 345 (2001).
\bibitem{wolf}
S. A. Wolf, et al., Science {\bf 294}, 1488 (2001)
\bibitem{casanova}
F. Casanova, et al., Phy. Rev. B. {\bf 79}, 184415 (2009)
\bibitem{chung}
S. B. Chung, et al., Phy. Rev. Lett. {\bf 121}, 167001 (2018) 
\bibitem{baibich}
M. N. Baibich, J. M. Broto, A. Fert, F. Nguyen Van Dau, and F. Petroff, 
Phys. Rev. Lett. {\bf 61}, 2472 (1988)
\bibitem{binasch}
G. Binasch, P. Gr\"{u}nberg, F. Saurenbach, and W. Zinn, 
Phys. Rev. B {\bf 39}, 4828(R) (1989)
\bibitem{parkin}
S. S. P. Parkin, Phys. Rev. Lett. {\bf 71}, 1641 (1993)
\bibitem{chappert}
C. Chappert et. al, Nature Mat., {\bf 6}, 147 (2007).
\bibitem{kadigrobov}
A. Kadigrobov, R. I. Shekhter and M. Jonson, Europhys. Lett. {\bf 54}, 394 (2001)
\bibitem{zutic}
I. \u{Z}uti\'{c}, J. Fabian and S. Das. Sarma, Rev. Mod. Phys. {\bf 76}, 323 (2004).
\bibitem{buzdin}
A.I. Buzdin, Rev. Mod. Phys. {\bf 77}, 935 (2005).
\bibitem{petrashov}
V. T. Petrashov, I. A. Sosnin, I. Cox, A. Parsons, and C. Troadec,
Phys. Rev. Lett. {\bf 83}, 3281 (1999)
\bibitem{flokstra}
M. G. Flokstra, et al., Nature (London) {\bf 12}, 57 (2016)
\bibitem{slonczewski}
J. C. Slonczewski, J. Magn. Magn. Mater, 159, L1 (1996)
\bibitem{berger}
L. Berger, Phys. Rev. B. 54, 9353 (1996)
\bibitem{halterman1}
K. Halterman, P. H. Barsic, and O. T. Valls, Phys. Rev. Lett. {\bf 99}, 127002 (2007)
\bibitem{halterman2}
K. Halterman, O. T. Valls and P. H. Barsic, Phys. Rev. B {\bf 77}, 174511 (2008)
\bibitem{golubov}
A. A. Golubov and M. Yu. Kupriyanov, Nature Materials {\bf 16}, 156 (2017)
\bibitem{zhu11}
Y. Zhu, A. Pal, M. G. Blamire and Z. H. Barber, Nature Materials {\bf 16}, 195 (2017)
\bibitem{bergeret}
F. S. Bergeret, A. F. Volkov and K. B. Efetov, Rev. Mod. Phys. {\bf 77}, 1321 (2005).
\bibitem{blamire}
M. G. Blamire and J. W. A. Robinson, J. Phys. Cond. Matter, {\bf 26},
453201, (2014)
\bibitem{eschrig1}
M. Eschrig, Phys. Today, {\bf 64}(1), 43 (2011)
\bibitem{eschrig2}
M. Eschrig, Rep. Prog. Phys. {\bf 78}(1), 104501(2015)


\bibitem{bardeen}
J. Bardeen, L. N. Cooper and J. R. Schrieffer, Phys. Rev. {\bf 108}, 1175 (1957).
\bibitem{saxena}
S.S. Saxena, et al., Nature (London) {\bf 406}, 587 (2005)
\bibitem{aoki}
D. Aoki, et al., Nature (London) {\bf 413}, 613 (2001).
\bibitem{pfleiderer}
C. Pfleiderer et al., Nature (London) {\bf 412}, 58 (2001).
\bibitem{huy}
N.T. Huy, et al., Phys. Rev. Lett. {\bf 99}, 067006 (2007).
\bibitem{bauer1}
E. Bauer, et al., Phys. Rev. Lett. {\bf 92}, 027003 (2004).
\bibitem{bauer2}
E. Bauer, I. Bonalde, and  M. Sigrist., Low Temp. Phys. {\bf 31}, 748 (2005).
\bibitem{bauer3}
E. Bauer, et al., J. Phys.Soc. Jpn. {\bf 76}, 051009 (2007).
\bibitem{motoyama}
G. Motoyama, et al., J. Phys. Conf. Ser. {\bf 400}, 022079 (2012).
\bibitem{kawasaki}
I. Kawasaki, et al., J. Phys. Soc. Jpn {\bf 82}, 084713 (2013).
\bibitem{akazawa}
T. Akazawa, et al., J. Phys. Cond. Matter {\bf 16}, L29 (2009).
\bibitem{anand1}
V. K. Anand, et al., Phys. Rev. B. {\bf 83}, 064522 (2011).
\bibitem{smidman}
M. Smidman, et al., Phys. Rev. B. {\bf 89}, 094509 (2014).
\bibitem{anand2}
V. K. Anand, et al., Phys. Rev. B. {\bf 90}, 041513 (2014).
\bibitem{matthias}
B. T. Matthias, V. B. Compton and E. Corenzwit, J. Phys. Chem. Solids {\bf 19}, 130 (1961).
\bibitem{singh}
R. P. Singh, et al., Phys. Rev. Lett. {\bf 112}, 107002 (2014).
\bibitem{pecharsky}
V. K. Pecharsky,  L. L. Miller and  K. A. Gschneidner, Phys. Rev. B {\bf 58}, 497 (1998).
\bibitem{hillier}
A. D. Hillier, J. Quintanilla and  R. Cywinski, Phys. Rev. Lett. {\bf 102}, 117007 (2009).
\bibitem{bonalde}
I. Bonalde, et al., New J. Phys. {\bf 13}, 123022 (2011).
\bibitem{yogi}
M. Yogi, et al., Phys. Rev. Lett. {\bf 93}, 027003 (2004).
\bibitem{ali}
M. N. Ali, et al., Phys. Rev. B {\bf 89}, 020505(R) (2014).
\bibitem{xu}
C. Q. Xu, et al., Phys. Rev. B {\bf 96}, 064528 (2017).
\bibitem{flouquet}
J. Flouquet and A. Buzdin, Phys. World {\bf 15}, 41 (2002).
\bibitem{nandi}
S. Nandi et al., Phys. Rev. B {\bf 89}, 014512 (2014).
\bibitem{samokhin}
K. V. Samokhin, E. S. Zijlstra and S. K. Bose, Phys. Rev. B. {\bf 69}, 094514 (2004).
\bibitem{sergienko}
I. A. Sergienko and S. H. Curnoe, Phys. Rev. B. {\bf 70}, 214510 (2004).
\bibitem{frigeri}
P. A. Frigeri, et al., Phys. Rev. Lett. {\bf 92}, 097001 (2004).
\bibitem{fujimoto1}
S. Fujimoto, Phys. Rev. B {\bf 72}, 024515 (2005).
\bibitem{fujimoto2}
S. Fujimoto, J. Phys. Soc. Jpn. {\bf 76}, 051008 (2007).
\bibitem{togano}
K. Togano, et al., Phys. Rev. Lett. {\bf 93}, 247004 (2004).
\bibitem{yuan}
 H. Q. Yuan, et al., Phys. Rev. Lett. {\bf 97}, 017006 (2006).
\bibitem{badica}
P. Badica, T. Kondo, and K. Togano, J. Phys. Soc. Jpn. {\bf 74}, 1014 (2005).


\bibitem{bychkov}
Y. A. Bychkov and E. I. Rashba,  J. Phys. C {\bf 17}, 6039 (1984)
\bibitem{molenkamp}
W. Molenkamp, G. Schimdt and G. E. W. Bruer, Phys. Rev. B {\bf 64}, 121202(R) (2001).
\bibitem{gorkov}
L. P. Gor'kov, E. I. Rashba, Phys. Rev. Lett. {\bf 87}, 037004 (2001).
\bibitem{bauer4}
E. Bauer, et al., Phys. Rev. B {\bf 80}, 064504 (2009).
\bibitem{wakui}
K. Wakui, et al., J. Phys. Soc. Jpn. {\bf 78}, 034710 (2009).
\bibitem{ribeiro}
R. L. Ribeiro, et al., J. Phys. Soc. Jpn. {\bf 78}, 115002 (2009).
\bibitem{biswas}
P. K. Biswas, et al., Phys. Rev. B {\bf 84}, 184529 (2011).
\bibitem{kuroiwa}
S. Kuroiwa, et al., Phys. Rev. Lett. {\bf 100}, 097002 (2008).
\bibitem{chen1}
J. Chen, et. al, Phys. Rev. B {\bf 83}, 144529 (2011).
\bibitem{chen2}
J. Chen, et al., New J. Phys. {\bf 15}, 053005 (2013).
\bibitem{wu}
S. Wu, K.V. Samokhin, Phys. Rev. B {\bf 82}, 184501 (2010). 

\bibitem{wu2}
C. T. Wu, O. T. Valls and K. Halterman, Phys. Rev. B. {\bf 90}, 054523 (2014)
\bibitem{romeo11}
F. Romeo. et al., Sci. Rep. 5, 17544 (2015)
\bibitem{shigeta}
I. Shigeta, et al., Appl. Phys. Lett. {\bf 112}, 072402 (2018)
\bibitem{beiranvand1}
R. Beiranvand, H. Hamzehpour, and M. Alidoust, Phys. Rev. B {\bf 94}, 125415 (2016)
\bibitem{beiranvand2}
R. Beiranvand, H. Hamzehpour, and M. Alidoust, Phys. Rev. B {\bf 96}, 161403(R) (2017)
\bibitem{halterman3}
K. Halterman,  O. T. Valls and M. Alidoust, Phys. Rev. Lett. {\bf 111}, 046602 (2013)
\bibitem{wu1}
S. Wu, K.V. Samokhin, Phys. Rev. B {\bf 81}, 214506 (2010).
\bibitem{hashimoto}
T. Hashimoto, A. A. Golubov, Y. Tanaka, and J. Linder, Phys. Rev. B {\bf 96}, 134508 (2017)

\bibitem{fominov1}
Ya. V. Fominov, N. M. Chtchelkatchev, and A. A. Golubov, Phys. Rev. B {\bf 66}, 014507 (2002)
\bibitem{fominov2}
Ya. V. Fominov, A. A. Golubov, and M. Yu. Kupriyanov, JETP Lett. {\bf 77}, 510 (2003)
\bibitem{fominov3}
Ya. V. Fominov, et al., JETP Lett. {\bf 91}, 308 (2010)
\bibitem{romeo}
F. Romeo and R. Citro, Phys. Rev. Lett., {\bf 111}, 226801 (2013)
\bibitem{olthof}
L. A. B. Olde Olthof, et al., Phys. Rev. B {\bf 98}, 014508 (2018).
\bibitem{tagirov}
L. R. Tagirov, Phys. Rev. Lett. {\bf 83}, 2058 (1999)
\bibitem{zhu}
J. Zhu and I. N. Krivorotov, Phys. Rev. Lett. {\bf 105} 207002 (2010)
\bibitem{alidoust1}
M. Alidoust, K. Halterman and O. T. Valls, Phys. Rev. B {\bf 92}, 014508 (2015)
\bibitem{alidoust2}
M. Alidoust, K. Halterman, Phys. Rev. B {\bf 97}, 064517 (2018)
\bibitem{halterman4}
K. Halterman, and M. Alidoust, Phys. Rev. B {\bf 94}, 064503 (2016)
\bibitem{acharjee}
S. Acharjee and U. D. Goswami, J. Appl. Phys. {\bf 120}, 263902 (2016).
\bibitem{gu}
J. Y. Gu, et al., Phys. Rev. Lett. {\bf 89}, 267001 (2002)
\bibitem{nowak}
G. Nowak, et al., Phys. Rev. B {\bf 78}, 134520 (2008)
\bibitem{leksin}
P. V. Leksin, et al., Phys. Rev. Lett. {\bf 109}, 057005 (2012)
\bibitem{zdravkov}
V. I. Zdravkov, et al., Phys. Rev. B {\bf 87}, 144507 (2013)
\bibitem{jara}
A. A. Jara, et al., Phys. Rev. B {\bf 89},184502 (2014)
\bibitem{khaydukov}
Yu. N. Khaydukov, et al., Phys. Rev. B {\bf 90}, 035130 (2014)
\bibitem{khaydukov1}
Yu. N. Khaydukov, et al., Phys. Rev. B {\bf 97}, 144511 (2018)
\bibitem{srivastava}
A. Srivastava, et al., Phys. Rev. Appl. {\bf 8}, 044008 (2017)
\bibitem{keizer}
R. S. Keizer, et al., Nature Lett. (London) {\bf 439}, 825 (2006)
\bibitem{kuerten}
L. Kuerten, et al., Phys. Rev. B {\bf 96}, 014513 (2017)

\bibitem{linder1}
J. Linder and A. Sudb{\o}, Phys. Rev. B {\bf 75}, 134509 (2007).
\bibitem{bozovic1}
M. Bo\u{z}ovi\'{c} and Z. Radovi\'{c}, Phys. Rev. B. {\bf 66}, 134524 (2002).
\bibitem{bozovic2}
M. Bo\u{z}ovi\'{c} and Z. Radovi\'{c}, New J. Phys. {\bf 9}, 264 (2007).
\bibitem{zutic1}
I. \u{Z}uti\'{c} and O. T. Valls, Phys. Rev. B {\bf 60}, 6320 (1999).
\bibitem{zutic2}
I. \u{Z}uti\'{c} and O. T. Valls, Phys. Rev. B {\bf 61}, 1555 (2000).
\bibitem{cheng}
Q. Cheng, D. Yu and B. Jin, Phys Lett. A {\bf 378}, 2900 (2014).
\bibitem{cheng1}
Q. Cheng, B. Jin, Physica B {\bf 426}, 40-44 (2013).
\bibitem{tanaka}
Y. Tanaka and S. Kashiwaya, Phys. Rev. Lett. {\bf 74}, 3541 (1995).
\bibitem{moen1}
E. Moen and O. T. Valls, Phys. Rev. B {\bf 98}, 104512 (2018).

\bibitem{linder}
J. Linder and A. Sudb{\o}, Phys. Rev. B {\bf 76}, 054511 (2007).
\bibitem{acharjee1}
S. Acharjee and U. D. Goswami, 
2019 Supercond. Sci. Technol. https://doi.org/10.1088/1361-6668/ab17ec
\bibitem{iniotakis}
C. Iniotakis, et al., Phys. Rev. B {\bf 76}, 012501 (2007).
\bibitem{kashiwaya}
S. Kashiwaya, Y. Tanaka, N. Yoshida, and M. R. Beasley, Phys. Rev. B {\bf 60}, 3572 (1999).



\bibitem{an}
Z. An, F. Q. Liu, Y. Lin and C. Liu, Sci. Rep., {\bf 2}, 388 (2012) 
\bibitem{sun}
Q. F. Sun and X. C, Xie, Phys. Rev. B {\bf 72}, 245305 (2005).
\bibitem{moen2}
E. Moen and O. T. Valls, Phys. Rev. B {\bf 97}, 174506 (2018).
\bibitem{linder5}
J. Linder and T. Yokoyama and A. Sudb{\o}, Phys. Rev. B {\bf 79}, 224504 (2009).
\bibitem{halterman6}
K. Halterman, M. Alidoust, Phys. Rev. B {\bf 94}, 064503 (2016).
\bibitem{halterman5}
K. Halterman, M. Alidoust, Supercond. Sci. Technol. {\bf 29}, 055007 (2016).

\bibitem{kopu}
J. Kopu, M. Eschrig, J. C. Cuevas, and M. Fogelstr\"{o}m, Phys. Rev. B {\bf 69}, 094501 (2004).
\bibitem{samokhvalov}
A. V. Samokhvalov, R. I. Shekhter and A. I. Buzdin, Sci. Rep., {\bf 4}, 5671 (2014).
\bibitem{blonder}
G. E. Blonder, M. Tinkham, T.M. Klapwijk, Phys. Rev. B {\bf 25}, 4515 (1982). 
\end{thebibliography}
\end{document}